\documentclass[trans]{IEEEtran}
\usepackage{latexsym}
\usepackage{comment}
\usepackage{graphicx}
\usepackage{amssymb,amsmath,amsfonts,amsthm}
\usepackage{algorithm,algorithmic}
\usepackage{subfigure}
\usepackage{xspace}
\usepackage[usenames,dvipsnames]{color}
\usepackage{color, colortbl}
\usepackage{verbatim}
\usepackage{url}
\usepackage{bm,dsfont}
\usepackage[T1]{fontenc} 
\usepackage{soul}
\usepackage[hidelinks]{hyperref}
\usepackage{xcolor}
\usepackage{tabu}
\usepackage{multirow}
\usepackage{booktabs}
\usepackage{cite}

\def\BibTeX{{\rm B\kern-.05em{\sc i\kern-.025em b}\kern-.08em T\kern-.1667em\lower.7ex\hbox{E}\kern-.125emX}}
\newenvironment{list4}{
  \begin{list}{$\bullet$}{%
      \setlength{\itemsep}{0.05cm}
      \setlength{\labelsep}{0.2cm}
      \setlength{\labelwidth}{0.3cm}
      \setlength{\parsep}{0in} 
      \setlength{\parskip}{0in}
      \setlength{\topsep}{0in} 
      \setlength{\partopsep}{0in}
      \setlength{\leftmargin}{0.17in}}}
      {\end{list}}

\definecolor{Gray}{gray}{0.9}
\definecolor{LightCyan}{rgb}{0.88,1,1}
\definecolor{beaublue}{rgb}{0.74, 0.83, 0.9}
\definecolor{lavender(web)}{rgb}{0.9, 0.9, 0.98}
\definecolor{mintcream}{rgb}{0.96, 1.0, 0.98}
\definecolor{skyblue}{rgb}{0.53, 0.81, 0.92}
\definecolor{whitesmoke}{rgb}{0.96, 0.96, 0.96}

\newcommand{\nocontentsline}[3]{}
\newcommand{\tocless}[2]{\bgroup\let\addcontentsline=\nocontentsline#1{#2}\egroup}

\begin{document}

\title{A Survey on UAV-Aided Maritime Communications: Deployment Considerations, Applications, and Future Challenges}

\author{Nikolaos Nomikos, \IEEEmembership{Senior Member, IEEE}, Panagiotis K. Gkonis,\\ Petros S. Bithas, \IEEEmembership{Senior Member, IEEE}, and Panagiotis Trakadas
\thanks{This work was supported in part by the Affordable5G Project funded by the European Commission through the Horizon 2020 and 5G-PPP Programs (www.affordable5g.eu/) under Grant H2020-ICT-2020-1 and Grant 957317.}
\thanks{N. Nomikos and P. Trakadas are with the Department of Port Management and Shipping, National and Kapodistrian University of Athens, 34400 Euboea, Greece. {\it Emails: {\tt nnomikos@ieee.org, ptrakadas@uoa.gr}.}}
\thanks{P. K. Gkonis and P. S. Bithas are with the Department of Digital Industry Technologies, National and Kapodistrian University of Athens, 34400 Euboea, Greece. {\it Emails: {\tt \{pgkonis, pbithas\}@uoa.gr}.}}
}

\maketitle

{\begin{abstract} Maritime activities represent a major domain of economic growth with several emerging maritime Internet of Things use cases, such as smart ports, autonomous navigation, and ocean monitoring systems. The major enabler for this exciting ecosystem is the provision of broadband, low-delay, and reliable wireless coverage to the ever-increasing number of vessels, buoys, platforms, sensors, and actuators. Towards this end, the integration of unmanned aerial vehicles (UAVs) in maritime communications introduces an aerial dimension to wireless connectivity going above and beyond current deployments, which are mainly relying on shore-based base stations with limited coverage and satellite links with high latency. Considering the potential of UAV-aided wireless communications, this survey presents the state-of-the-art in UAV-aided maritime communications, which, in general, are based on both conventional optimization and machine-learning-aided approaches. More specifically, relevant UAV-based network architectures are discussed together with the role of their building blocks. Then, physical-layer, resource management, and cloud/edge computing and caching UAV-aided solutions in maritime environments are discussed and grouped based on their performance targets. Moreover, as UAVs are characterized by flexible deployment with high re-positioning capabilities, studies on UAV trajectory optimization for maritime applications are thoroughly discussed. In addition, aiming at shedding light on the current status of real-world deployments, experimental studies on UAV-aided maritime communications are presented and implementation details are given. Finally, several important open issues in the area of UAV-aided maritime communications are given, related to the integration of sixth generation (6G) advancements. These future challenges include physical-layer aspects, non-orthogonal multiple access schemes, radical learning paradigms for swarms of UAVs and unmanned surface and underwater vehicles, as well as UAV-aided edge computing and caching. \end{abstract}

}

\section{Introduction}\label{intro}
In the past decades, wireless networks have been evolving towards supporting users services, which are located in urban environments, while the fourth and fifth generations (4G and 5G) of mobile communications have put special emphasis in the coexistence of mobile users and Internet-of-Things (IoT) devices \cite{javaid2018cm,sharm2020comst,jian2021ojcoms,vaezi2022comst}. Unfortunately, most network architectures and communication techniques were designed for land-based communications, while the maritime domain has been largely neglected from this revolution. As a result it is mainly based on the satellite segment, with the known issues of high-latency and low-data rates \cite{xia2020wcm}. Considering that the vast majority of trade relies on maritime transportation, while the interest for a wide range of maritime activities, such as ocean exploration for natural resources and pollution monitoring, has spiked, a radical shift to maritime communications is needed \cite{wang2020machine}. In this context, the rise of unmanned aerial vehicles (UAVs) and their integration in wireless networks provides a viable means for achieving broadband coverage at sea \cite{zeng2016cm, Wang2021jsac, bithas2019sensors}. UAVs are capable of flexibly providing radio-resources depending on the Quality-of-Service (QoS) requirements of applications and can operate autonomously in a distributed manner, ensuring high reliability and low-latency. 

Thus, it is critical to efficiently integrate UAVs in maritime environments in order to achieve ubiquitous connectivity  and support maritime applications on the water surface and underwater \cite{jahanbakht2021comst}. In this survey, the role of UAVs in facilitating maritime communication networks (MCNs) is presented and current solutions are given in detail. For this purpose, relevant communication techniques are discussed and categorized according to network layer and performance target. Moreover, several network architecture propositions, comprising UAVs, satellites, terrestrial and heterogeneous maritime network nodes are shown and various open issues stemming from this important research area are given in detail, targeting to stimulate further interest in UAV-aided maritime communications.

\subsection{Maritime communications}\label{mar_comm}

MCNs aim at supporting applications related among others, to trade, ocean exploration, pollution monitoring, marine tourism and search and rescue (SAR) operations \cite{guan2021cm, haidine2021agers}. Such an ecosystem relies on a heterogeneous mix of vessels, buoys, platforms, unmanned surface vehicles (USVs) and unmanned underwater vehicles (UUVs), sensors, and actuators \cite{Wang2021jsac}. In addition, maritime services are characterized by varying QoS types, since, for example, crew and passengers on-board cruise ships might be interested in broadband connectivity, SAR operations entail the transmission of real-time video, while IoT services in smart maritime environments and intelligent transportation systems are based on Ultra-reliable and ultra-low latency (URLLC) \cite{alqurashi2022arxiv, matracia2022ojcoms}. 
\begin{table*}[t]
\caption{List of surveys presenting MCN solutions.}
\label{surveys_comp}
\centering
{\begin{tabular}[t]{m{10em}<{\raggedright} m{14em}<{\raggedright} m{34em}<{\raggedright}}
\hline
\rowcolor{whitesmoke}\vspace{0.12cm}\textbf{\hspace{0.8cm}Reference\vspace{0.12cm}}	& \vspace{0.12cm}\textbf{\hspace{1cm}Short summary\vspace{0.12cm}}  & \vspace{0.12cm}\textbf{\hspace{4.3cm}Contributions}\vspace{0.12cm}\\\hline
\rowcolor{whitesmoke}\rowcolor{beaublue}
\vspace{0.05cm}Wei et al.~\cite{wei2021iotj} & \vspace{0.05cm}A survey on hybrid satellite-terrestrial networks for maritime IoT applications & \vspace{0.05cm}
\begin{itemize} 
\item[-] Maritime communications challenges, related to meteorological and geographical characteristics and heterogeneous service requirements are discussed
\item[-] Hybrid-satellite terrestrial technologies for high transmission efficiency, wide network coverage, and maritime-specific services are discussed
\item[-] Very few works related to UAV-aided maritime communications are included\vspace{-0.2cm}
\end{itemize}\\
\rowcolor{whitesmoke}\vspace{0.05cm}Jahanbakht et al.~\cite{jahanbakht2021comst} &\vspace{0.05cm} A survey on the synergy among the Internet of Underwater Things (IoUT) and big marine data analytics
&\vspace{0.05cm} 
\begin{itemize}
\item[-]A thorough overview of IoUT network architectures, communication techniques and state-of-the-art research challenges is presented
\item[-]The role of big data analytics and machine learning for IoUT applications is given and relevant frameworks and platforms are discussed
\item[-]UAV-based solutions and related aspects are missing from the survey and the focus is on underwater communications and applications \vspace{-0.2cm}
\end{itemize}\\%
\rowcolor{beaublue}
\vspace{0.05cm}Alqurashi et al.~\cite{alqurashi2022arxiv} &\vspace{0.05cm} A survey on enabling
technologies, opportunities, and challenges of maritime communications &\vspace{0.05cm} 
\begin{itemize}
\item[-]Focus on physical-layer techniques, channel modelling and RRM  for maritime communications
\item[-]Presentation of emerging MCN use cases and open challenges related to THz and visible light communications and data-driven optimization
\item[-]Very few works on UAV-aided maritime communications are included \vspace{-0.2cm}
\end{itemize}\\%
\rowcolor{whitesmoke}\vspace{0.05cm}This survey &\vspace{0.05cm} A survey emphasizing on the integration of UAVs for supporting the operation of maritime communication networks &\vspace{0.05cm}
\begin{itemize}
\item[-]UAV-aided maritime network architectures and the role of the heterogeneous network nodes are presented
\item[-]Depending on the network layer that the communication techniques operate, UAV-aided MCN solutions are categorized and critically evaluated
\item[-]UAV-aided maritime communications open issues, ranging from the physical-layer up to cloud/edge architectures and ML integration are discussed \vspace{-0.2cm}
\end{itemize}\\%
\hline
\end{tabular}}
\end{table*}

The current paradigm for providing coverage to maritime activities depends on shore-based base stations (BSs) and satellite constellations. Unfortunately, such maritime networking architectures are not capable of supporting the emerging maritime applications, due to low data rates, high communication delays, and unreliable connectivity. Although there have been industrial initiatives on both terrestrial and satellite segments, with broadband satellite coverage and long-distance shore-to-vessel communications using cellular standards \cite{starlink, hyoungwon2017ict, huo2020iotj}, further research on developing flexible and intelligent maritime networks is required. Towards this end, exploiting aerial nodes, as envisioned in various 6G network architectures can mitigate the impact of geographical characteristics on path-loss, allow reduced communication delays, and enhance the communications reliability through additional wireless paths for data transmissions.

\subsection{UAV-aided wireless networks}\label{UAV_comm}

6G networks are envisioned to support novel network architecture paradigms, encompassing moving nodes for dynamic resource provisioning and increased network resiliency \cite{bithas2020uav,9040264}. In this context, the integration of UAVs to complement terrestrial and satellite networks has been investigated in various recent works \cite{zhu2021iotj, Li2020wc, Wang2021jsac}. UAV-aided networks offer extended coverage in remote and rural settings, quick recovery after disasters and emergency situations,  ubiquitous communications in flash crowd traffic demands, enabling various IoT use cases, such as precision agriculture and fleet management \cite{zhao2019vtm, matracia2022ojcoms}.  

In maritime settings, the use of UAVs for satisfying heterogeneous maritime services, requiring broadband connectivity or URLLC for IoT has attracted significant interest in recent years \cite{zolich2019irsj}. More specifically, UAVs can be flexibly deployed and assume the role of wireless relays, enabling multi-hop communication between ground BSs and sea vessels. In the case of underwater IoT, UAVs can greatly facilitate data collection by cooperating with USVs and UUVs, enhancing the capabilities of ocean monitoring systems \cite{luo2022sensorsj}. Meanwhile, during SAR operations, UAVs can provide high capacity line-of-sight (LoS) links in order to facilitate the transmission of real-time video data among the participating vessels and ground stations \cite{wang2018access}. An inherent characteristic of UAVs is that energy-efficient communications and trajectory optimization algorithms have to be put in place, while recent advancements in wireless power transfer (WPT) can offer further gains towards extending their flight time.

\begin{figure*}[t]
\centering
\includegraphics[width=\textwidth]{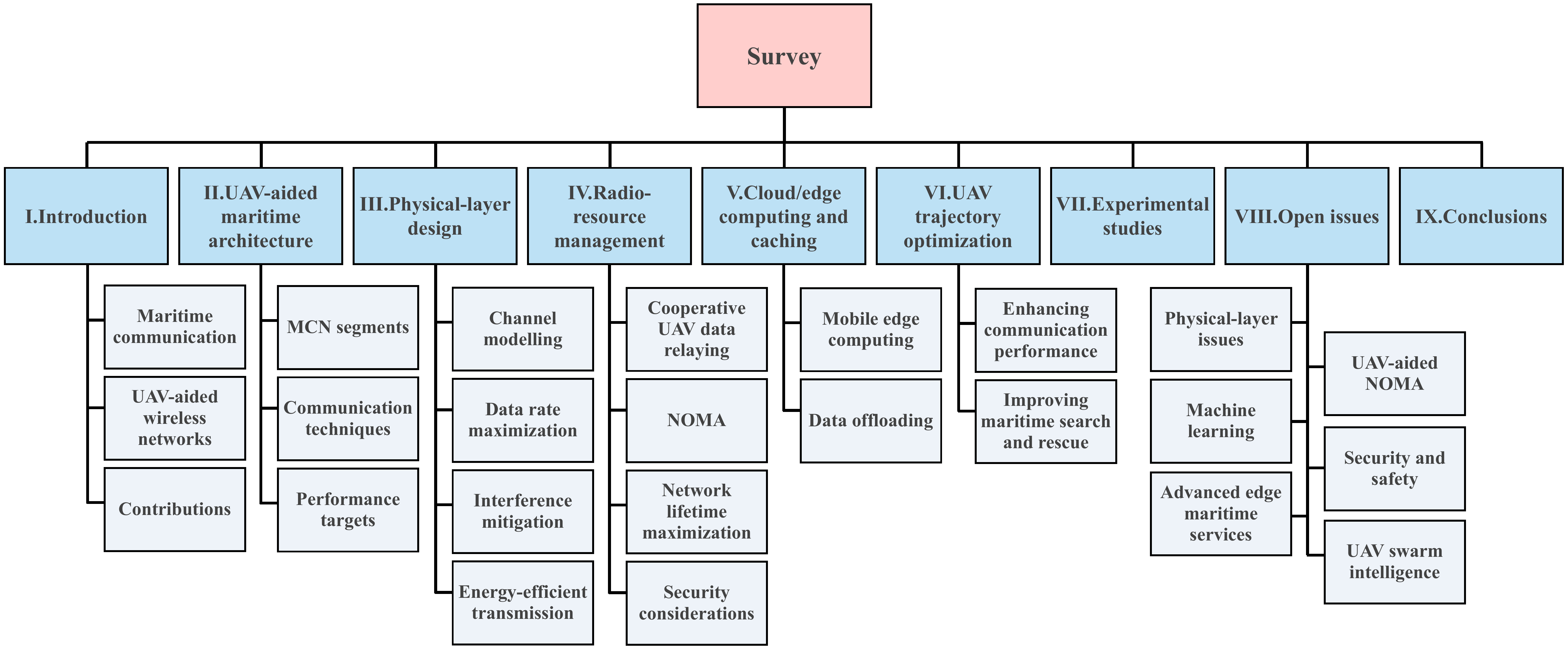}
\caption{Survey structure.}
\label{structure}
\end{figure*}

\subsection{Contributions}\label{contr}
In recent years, the interest in maritime activities has increased and currently deployed communication infrastructure cannot cope with the requirements of emerging use cases \cite{wei2021iotj, alqurashi2022arxiv}. In this context, UAV-aided solutions represent a radical paradigm shift that complements terrestrial and satellite segments, bringing several unique advantages in terms of deployment flexibility, as well as path-loss and delay reduction. Taking into consideration the potential of integrating UAV in maritime networks, this survey provides a thorough overview of relevant solutions and categorizes them according to the network layer issues and performance target that they address. In greater detail, our contributions are as follows:

\begin{list4}
    \item The current status in maritime network architectures is presented, and the integration and role of UAVs is discussed. Also, details on the relevant building blocks are given.
    \item The physical-layer, resource management, and cloud/edge UAV-aided algorithms for maritime communications are categorized based on their performance targets. Also, recent advancements in UAV trajectory optimization for maritime applications are thoroughly discussed.
    \item Then, aiming at shedding light on the current status of real-world deployments, experimental studies on UAV-aided maritime communications are presented and implementation details are given. 
    \item Several open issues are highlighted, including intelligent reflecting surface (IRS)-aided communications, non-orthogonal multiple access (NOMA), swarm intelligence for UAVs, USVs and UUVs, and UAV flight time maximization through WPT.
\end{list4}

Table~{\ref{surveys_comp}} summarizes the contributions of surveys in the field of maritime communications and maritime IoT and their emphasis on UAV integration issues is highlighted. Starting with the survey by Wei {\it et al.} \cite{wei2021iotj}, the authors focus on hybrid-satellite network architectures towards supporting the requirements of maritime IoT. Details on maritime communications challenges are discussed, related to environmental and geographical characteristics, and state-of-the-art solution are categorized according to their goal, i.e., improving the transmission efficiency, ensuring wide coverage and guaranteeing maritime service QoS. However, this survey only includes very few studies on UAV-aided maritime communications. Then, the survey by Jahanbakht {\it et al.} \cite{jahanbakht2021comst} provides a comprehensive overview and tutorial on the synthesis of IoUT and big marine data analytics. Several IoUT architectures are presented, and details on the interplay among IoUT and machine-learning-aided optimization are provided. However, this survey significantly differs to the one presented in this paper, as the state-of-the-art in UAV-aided maritime networks is completely absent. Another recent survey by Alqurashi {\it et al.} gives an overview of maritime communications issues and focuses on physical-layer aspects, current developments in channel models and radio-resource management (RRM) algorithms. Moreover, interesting maritime use cases are discussed, highlighting the research potential in this area. Nonetheless, UAV-based solutions are not discussed in the survey, thus leaving a gap in the literature that this work aims to fill.

\subsection{Structure}\label{str}
The structure of this survey is as follows. First, Section~\ref{architecture} presents the state-of-the-art in architecture designs for UAV-aided maritime network and provides details on their components. Then, Section~\ref{PHY} includes works focusing on physical-layer issues of UAV-aided maritime communications. Resource management and multiple access aspects are discussed in Section~\ref{RRM}, while Section~\ref{cloud_edge} focuses on the integration of cloud/edge computing and caching for improving the performance of UAV-aided maritime applications. Subsequently, optimal UAV trajectory design in maritime environments is the topic of Section~\ref{trajectory} and experimental implementations of UAV-aided maritime topologies are discussed in Section~\ref{experimental}. Moreover, several important open issues in the area of UAV-aided maritime communications are given in Section~\ref{open_issues}. Finally, conclusions are given in Section~\ref{conclusions}. Overall, the structure of this survey is shown in Fig.~\ref{structure}, while Table~II includes the list of acronyms being used throughout this survey.

\begin{table*}[ht]
\renewcommand{\arraystretch}{1.4}
{
\centering
\caption{List of acronyms }
\label{abbreviations}
\begin{tabular}[t]{  m{3.2em} m{15 em} m{3.2em}  m{15em} m{3.2em} m{15 em} }
  \hline\rowcolor{whitesmoke}
  \textbf{Acronym} & \hspace{1.5cm}\textbf{Definition} & \textbf{Acronym} & \hspace{1.5cm}\textbf{Definition} & \textbf{Acronym} & \hspace{1.5cm}\textbf{Definition} \\\hline
    \rowcolor{beaublue}BER & Bit error rate  & BS &  Base station & CCI & co-channel interference   \\
    \rowcolor{whitesmoke} DCGAN & Deep convolutional generative adversarial network & D2D & Device-to-device  & DDPG & Deep deterministic policy gradient \\
    \rowcolor{beaublue} DRL & Deep reinforcement learning & EE &  Energy efficiency & FDTD &  Finite difference domain method \\
    \rowcolor{whitesmoke} FL & Federated learning & FSO & Free space optics & GEO & Geostationary earth orbit \\
    \rowcolor{beaublue} HAP & High altitude platforms & IoT & Internet-of-Things & IoUT & Internet of underwater things \\
    \rowcolor{whitesmoke}IRS & Intelligent reflecting surfaces & LEO & Light emitting diode & LEO & Low Earth orbit \\
    \rowcolor{beaublue}LMS & Least mean square & LoS & Line of sight & LSTM & Long-short-term memory \\
    \rowcolor{whitesmoke} LTE & Long term evolution & MCN & Maritime communication network & MDP & Markov decision process \\
    \rowcolor{beaublue} MEC & Mobile edge computing  & MIMO & Multiple-input multiple-output & mMIMO & Massive MIMO \\
    \rowcolor{whitesmoke}ML & Machine learning & NLoS & Non-line-of-sight  & NOMA & Non-orthogonal multiple access \\
    \rowcolor{beaublue}OMA & Orthogonal multiple access & OMN & Ocean monitoring & PLS & Physical-layer security \\
    \rowcolor{whitesmoke}PSO & Particle swarm optimization &  QoS & Quality-of-Service  & RAN & Radio access network \\
    \rowcolor{beaublue}RF & Radio frequency & RL & Reinforcement learning & RRM & Radio resource management \\ 
    \rowcolor{whitesmoke}SAR & Search and rescue & SCA & Successive convex optimization & SE & Spectral efficiency\\
    \rowcolor{beaublue} SN & Sink node & SNR & Signal-to-noise ratio & TBS & Terrestrial base station \\
    \rowcolor{whitesmoke}TDMA & Time division multiple access & UAV & Unmanned aerial vehicle & USN & Underwater sensor network \\
    \rowcolor{beaublue}USV & Unmanned surface vehicle & UUV & Unmanned underwater vehicles &URLLC & Ultra-reliable and ultra-low latency  \\
    \rowcolor{whitesmoke} VHF & Very high frequency & VLC & Visible light communication & WPT & Wireless power transfer  \\
  \hline
\end{tabular}
}
\label{notation}
\end{table*} 

\begin{figure*}[t]
\centering
\includegraphics[width=0.85\textwidth]{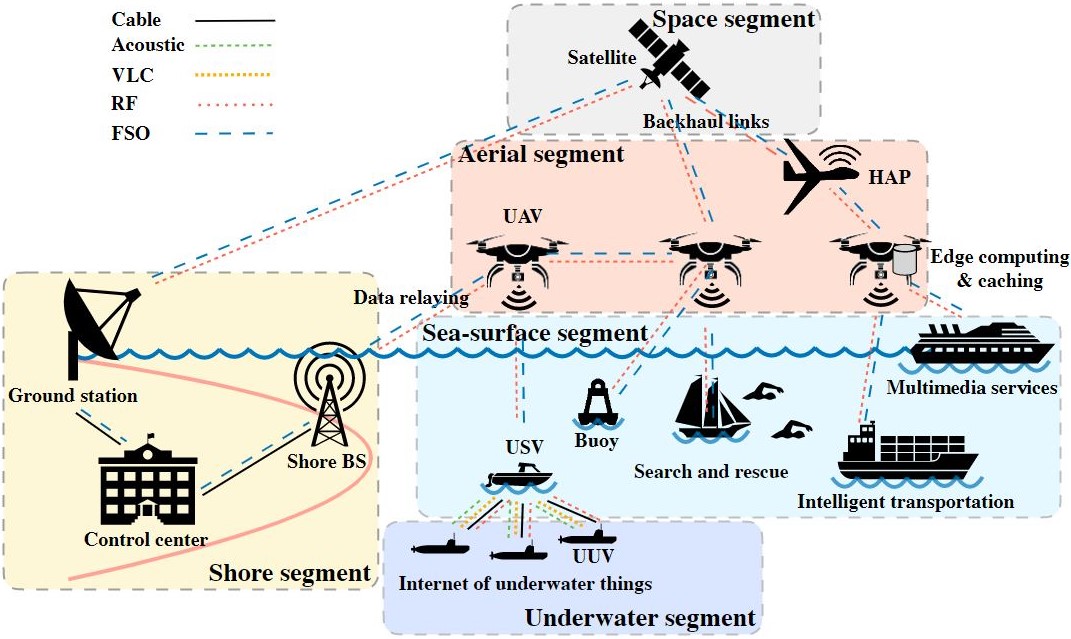}
\caption{An illustrative UAV-aided maritime communications architecture.}
\label{maritime_arch}
\end{figure*}

\section{UAV-Aided Maritime Network Architecture}\label{architecture}

Maritime activities rely on heterogeneous network topologies, where numerous UUVs, USVs, sea vessels, buoys, platforms, and sensors cooperate with UAVs and satellites to enable reliable communications in highly mobile and volatile environments. An illustrative UAV-aided maritime communications architecture is depicted in Fig.~\ref{maritime_arch}. Below, the main architectural segments presented in that figure will be analyzed, as well as the different communication techniques adopted and the various performance targets assumed.

\subsection{MCN segments}
As it can be observed in Fig.~\ref{maritime_arch}, four main segments can be identified in an MCN architecture. More specifically, a maritime segment involving underwater and surface activities, a shore segment, an aerial segment and a space segment \cite{ai2020wcm}.

\subsubsection{Maritime segment}
Here, the edge nodes of the MCN are located, being responsible for surface and underwater data acquisition, cooperative relaying with multiple modes of communications, and edge computing tasks.

\textbf{Underwater:} In an MCN, the underwater segment includes sensor nodes, underwater buoys and UUVs, being responsible for marine data acquisition and exchange that is forwarded to other MCN nodes, such as UUVs, ships or UAVs. As electromagnetic waves are subjected to high attenuation in seawater when data must be transmitted to longer distances, the nodes of the underwater segment rely on acoustic signals.

\textbf{Sea surface:} On the sea surface, ships, USVs, and buoys reside, communicating to support applications related, among others, to intelligent transportation, environmental observation, data relaying from/to the underwater segment, and maritime SAR. Recently, the introduction of extended coverage giant cells, in the form of seaborne floating towers, being semi-submersible steel reinforced concrete platforms has been proposed \cite{guan2021cm}.

\subsubsection{Shore segment}
The shore segment hosts BSs that provide coverage to nearby maritime nodes and UAVs, employing cellular standards. Furthermore, on the ground stations communicating with the space segment are present, allowing the transmission of data within the broader MCN.

\subsubsection{Aerial segment}
The existence of shore BS and satellites provides wide coverage to maritime activities but is insufficient to support the QoS of service types that require broadband connectivity or URLLC \cite{li2021tii}. Towards this end, UAV-aided MCNs have emerged where a flexible aerial segment allows the dynamic provisioning of radio-resources to remote areas, low-latency compared to satellite links, and high reliability with multi-hop transmissions and diverse wireless paths \cite{aitallal2022ai2sd}. Fig.~\ref{maritime_arch} depicts various maritime applications that are supported by UAVs, including data relaying to/from the underwater and surface segments for maritime IoT purposed, support for maritime SAR and communication with sea vessels desiring intelligent transportation and multimedia service provisioning. In addition, to UAVs, the aerial segment can employ high-altitude platforms (HAPs) mainly residing at the stratosphere.

\subsubsection{Space segment}
The space segment encompasses different satellite systems, in the sense of geostationary earth orbit (GEO)-based INMARSAT and low-earth orbit (LEO) constellations, such as Starlink. The main responsibilities of the space segment include its use as a back-up when the shore BSs and the aerial segment fail to provide coverage to maritime nodes and backhauling/fronthauling in order to enable data availability across the whole MCN structure. 

\subsection{Communication Technologies}
Stemming from the heterogeneity of MCN segments, different communication technologies are usually employed in order for the MCN to adjust to the various environmental and propagation characteristics. The vast majority of communication solutions are wireless and only in some underwater topologies, communication relies on wired connections. As a result, below we focus on the various wireless communication technologies for MCNs.

\subsubsection{Radio frequency (RF) communications}
In conventional maritime communication systems, RF-based transmissions operate on the very high frequency band (VHF) between 156--174 MHz and provide radio services and SAR support. Nonetheless, the provision of broadband services and maritime IoT applications will use cellular and Wi-Fi bands that can cope with high data rate and low-latency requirements, exploiting the re-position capabilities of UAVs for improved wireless connectivity \cite{zhong2019imm, teixera2020access, park2021milcom} and direct communication among maritime terminals \cite{liu2019tcc}. Also, the adoption of novel technologies, such as IRS can mitigate the degrading effects of fading path-loss \cite{ramezani2022network}.

\subsubsection{Free space optical (FSO) communications}
The constantly increasing desire for high-throughput services and the spectrum crunch that is experienced in the RF band has motivated researchers to introduce alternative communication paradigms. A popular solution for high bandwidth transmissions is FSO communication guaranteeing large bandwidth, unlicensed spectrum, high data rate, fast deployment and power reduction, under LoS conditions \cite{kaushal2017comst}. However, FSO transmissions are affected by atmospheric effects, i.e.,  absorption, scattering, and turbulence \cite{lionis2021optics}. As a result, various works have studied hybrid RF/FSO solutions that leverage the advantages of both communication technologies, switching to the most appropriate one when environmental characteristics dynamically change, addressing misalignment, resource allocation and energy efficiency issues \cite{jamali2016twc, arienzo2019tccn, upadhya2020tcom, ninos2021ojcoms}.

\subsubsection{Visible-light communications (VLC)}

Another technology that has received several contributions in recent years is VLC, using light-emitting diodes (LEDs) to transmit data and avoid impairments of RF communications, such as interference and signal leakage. In maritime environments, VLC can especially facilitate communications of the underwater segment, among UUVs and from UUVs to USVs/ships/UAVs acting as data sinks but also between nodes above the sea surface \cite{huang2020jlt}. In this context, efficient VLC solutions should consider the mobility of maritime IoT deployments and ensure accurate pointing between the VLC transceivers \cite{eso2021ptl}.

\subsubsection{Acoustic communications}
In underwater communications, long-range transmission can be performed when acoustic waves are employed. In several studies, the challenges of underwater acoustic communications have been highlighted, including poor quality and highly dynamic nature of acoustic channels, smaller channel capacity compared to RF channels, and larger propagation delay \cite{jiang2018comst1, jiang2018comst2, luo2021comst}. Other works have proposed hybrid schemes where acoustic signals are used together with RF and optical signals to combat attenuation effects \cite{luo2019access}.

\subsection{Performance targets}
In order to decide the optimal mix of communication technologies and architectural segments, a wide range of MCN performance targets exists and in many cases introduces trade-offs among conflicting targets and implementation complexity. As it will be presented in the following sections, the various network procedures can be optimized either through conventional optimization techniques and/or machine learning (ML) algorithms \cite{yang2020network}. 

\subsubsection{Spectral efficiency}
An important metric in wireless communication systems is spectral efficiency, given in bps/Hz, representing the achieved data rate over the available bandwidth. Considering that the number of maritime networks' nodes is increasing, while in some areas, MCNs overlap with terrestrial networks, spectral resources become more scarce and a higher frequency re-use is needed, e.g., by deploying aerial BSs or resorting to other frequency bands. Meanwhile, the heterogeneity of MCNs, comprising satellite-aided backhauling calls for intelligent link selection solutions in order to improve the MCNs' spectral efficiency.

\subsubsection{Energy efficiency/network lifetime maximization}
As maritime activities are constantly evolving, an increased number of battery-dependent network nodes, such as UUVs, USVs, UAVs, is deployed. As a result, it is  vital to adopt energy-efficient communication techniques with power control \cite{nomikos2022ojcoms} and appropriate routing algorithms \cite{mukherjee2020tsc}. The energy efficiency is usually measured in bits/joule, i.e., the number of bits that are transmitted over the energy used for their transmission. In addition, sustainable operation with reduced carbon footprint and operational expenditure for the infrastructure and vessel owners is necessitated. Apart from capitalizing recent advancements in battery technology, WPT is a promising solution to prolong network lifetime of the different MCN segments \cite{hu2021tvt,muhammad2021tvt}.  

\subsubsection{Communication delay minimization}
As URLLC services are highly desirable in the context of critical maritime services, the end-to-end delay is a critical QoS parameter that determines the MCN performance. The reduction of end-to-end delay can be achieved by selecting high throughput wireless data routes through optimal network node selection \cite{nomikos2020vehcom}, starting from the physical-layer, considering parameters such as channel quality and the use of multiple antennas for increased diversity, and reaching up to the multiple access layer with non-orthogonal and grant-free schemes.

\subsubsection{Task and data offloading}
Under the mobile edge computing (MEC), edge nodes can provide their computing resources towards bringing data computation closer to the data acquisition points. In this way, latency is dramatically reduced, as long as efficient task allocation algorithms are utilized. In the context of data offloading, several studies aim to improve the cache hit ratio of edge caching-aided networks, defined as the ratio of cached files being requested by the end users over the total number of files that are stored in the cache. Thus, a high cache hit ratio corresponds to higher QoS and backhaul/fronthaul offloading, thus avoiding data fetching from/to remote servers when cache-aided nodes are deployed in the MCN \cite{feng2022ciot}.

\section{Physical-Layer Design}\label{PHY}

In this section various research efforts will be analyzed that deal with physical layer issues in the design and deployment of UAV-aided MCNs. In this context, major design goals that should be considered in parallel with the provision of acceptable QoS to end users include proper channel modelling for 5G/6G enabled communications that take into consideration the harsh maritime environment, and data rate maximization along with energy efficient transmissions, in order to deal with the limited battery life of UAVs.


\begin{table}[t]
\caption{List of papers on PHY-layer aspects of UAV-aided maritime communications.}
\label{PHY_table}
\centering
{\begin{tabular}[t]{m{7em}<{\raggedright} m{4.4em}<{\raggedright} m{6.9em}<{\raggedright} m{6.7em}<{\raggedright}}
\hline
\rowcolor{whitesmoke}\textbf{Reference}	& \textbf{Maritime topology} & \textbf{Communication target}  & \textbf{Method}\\
\hline%

\rowcolor{beaublue}\vspace{0.1cm} Timmins et al.~\cite{Timmins2009VEH} &\vspace{0.1cm} UAV-aided &\vspace{0.1cm} Channel modelling &\vspace{0.1cm} FDTD \\%

\rowcolor{whitesmoke}\vspace{0.1cm} Liu et al.~\cite{Liu2021JSAC} &\vspace{0.1cm} UAV-aided &\vspace{0.1cm} Channel modelling &\vspace{0.1cm} Multi-bounce components \\%

\rowcolor{beaublue}\vspace{0.1cm} Gao et al.~\cite{Gao2020China} &\vspace{0.1cm} UAV-aided &\vspace{0.1cm} Channel modelling &\vspace{0.1cm} Ray-tracing algorithm \\%

\rowcolor{whitesmoke}\vspace{0.1cm} Rasheed et al.~\cite{Rasheed2022ieee_trans} &\vspace{0.1cm} UAV-aided &\vspace{0.1cm} Channel modelling &\vspace{0.1cm} LSTM-DCGAN \\%

\rowcolor{beaublue}\vspace{0.1cm} Liu et al.~\cite{Liu2021JCN} &\vspace{0.1cm} UAV-aided &\vspace{0.1cm} Link efficiency &\vspace{0.1cm} Incorporation of weather measurements \\%

\rowcolor{whitesmoke}\vspace{0.1cm}Cao et al.~\cite{Cao2021} &\vspace{0.1cm} UAV-aided &\vspace{0.1cm} SE maximization &\vspace{0.1cm} Normalized LMS for transceiver optimization \\

\rowcolor{beaublue}\vspace{0.1cm}Wang et al.~\cite{Wang2020} &\vspace{0.1cm} Satellite-UAV-terrestrial &\vspace{0.1cm} Minimum user rate maximization &\vspace{0.1cm} Random matrix theory \\

\rowcolor{whitesmoke}\vspace{0.1cm}Ghanbari et al.~\cite{Ghanbari2022} &\vspace{0.1cm} UAV-UUV-aided &\vspace{0.1cm} Outage probability &\vspace{0.1cm} Theoretical model and simulations \\

\rowcolor{beaublue}\vspace{0.1cm}Wanq et al.~\cite{Wang2020ieee_internet} &\vspace{0.1cm} UAV-aided &\vspace{0.1cm} Path connectivity &\vspace{0.1cm} Theoretical model and simulations \\

\rowcolor{whitesmoke}\vspace{0.1cm}Fang et al.~\cite{Fang2020} &\vspace{0.1cm} Maritime cognitive satellite-UAV-terrestrial &\vspace{0.1cm} CCI mitigation &\vspace{0.1cm} Random matrix theory \\

\rowcolor{beaublue}\vspace{0.1cm}Liu et al.~\cite{Liu2021conf} &\vspace{0.1cm} UAV-aided &\vspace{0.1cm} Jamming mitigation &\vspace{0.1cm} DRL \\

\rowcolor{whitesmoke}\vspace{0.1cm}Rahimi et al.~\cite{Rahimi2022} &\vspace{0.1cm} UAV-aided &\vspace{0.1cm} Energy efficiency maximization &\vspace{0.1cm} Genetic algorithm \\

\rowcolor{beaublue}\vspace{0.1cm}Wang et al.~\cite{Wang2021jsac} &\vspace{0.1cm} Satellite-UAV-aided &\vspace{0.1cm} Minimize energy consumption &\vspace{0.1cm} Min-Max transformation \\

\rowcolor{whitesmoke}\vspace{0.1cm}Li et al.~\cite{Li2020ieeetrans} &\vspace{0.1cm} Satellite-UAV-terrestrial &\vspace{0.1cm} Data rate maximization &\vspace{0.1cm} successive convex optimization \\

\rowcolor{beaublue}\vspace{0.1cm}Hong et al.~\cite{Hong2021ieeenetwork} &\vspace{0.1cm} Space-air-ground-sea integrated network &\vspace{0.1cm} Energy efficiency maximization &\vspace{0.1cm} Shape-adaptive antennas \\

\rowcolor{whitesmoke}\vspace{0.1cm}Li et al.~\cite{Li2020wc} &\vspace{0.1cm} Satellite-UAV-terrestrial &\vspace{0.1cm} CCI mitigation &\vspace{0.1cm} successive
convex optimization \\

\hline
\end{tabular}}
\end{table}

\subsection{Channel modelling for maritime environments}

Unlike terrestrial broadband wireless networks, where channel modelling is mainly dictated by a large number of non-line-of-sight (NLoS) components, in a maritime wireless network there can be a large number of direct signal paths. Moreover, sea volatility and extreme weather conditions, especially in the oceans, can have a sever impact on channel conditions. Therefore, the wireless propagation model of the ocean is significantly different from the corresponding terrestrial one.

\begin{figure}[t]
\centering
\includegraphics[width=\columnwidth]{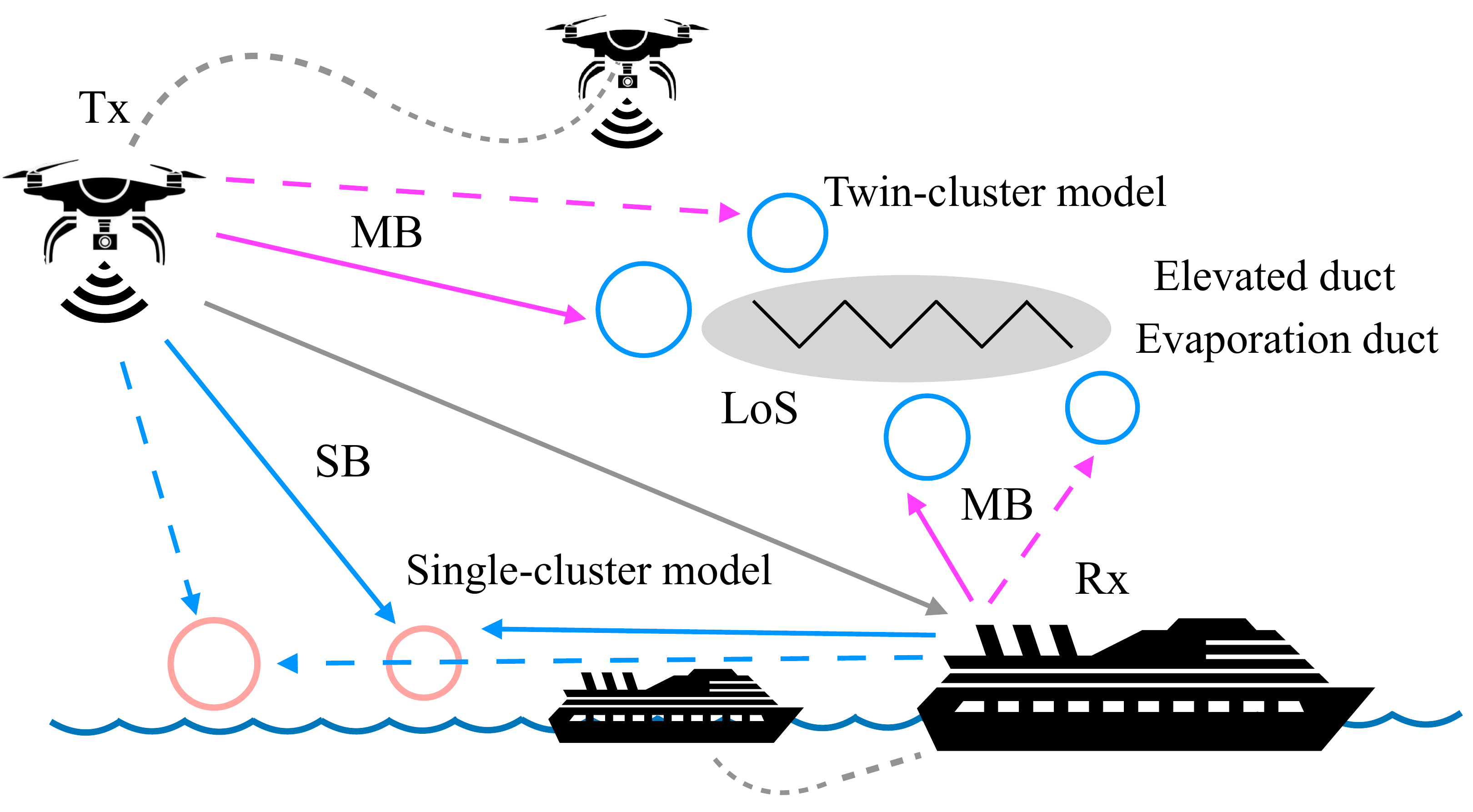}
\caption{Elements of the non-stationary multi-mobility UAV-to-ship channel model \cite{Liu2021JSAC}.}
\label{channel_model}
\end{figure}

In \cite{Timmins2009VEH}, the channel in a UAV-aided MCN was modelled with the help of the finite-difference time domain method (FDTD) method, in an effort to take into consideration the effects of sea surface shadowing conditions on the marine communications channel. Results were restricted for fixed Tx and Rx heights and frequencies below 3GHz. 
In \cite{Liu2021JSAC}, a novel non-stationary multi-mobility UAV-to-ship channel model is proposed. Fig.~\ref{channel_model} provides details of the non-stationary channel model and to this end, three major components have been included: the line-of-sight (LoS) component, the single-bounce components resulting from the fluctuation of sea water, and multi-bounce components introduced by the waveguide effect over the sea surface. This model supports multimobility as well, in the sense that vessels, UAVs, clusters, etc. can move to arbitrary locations.
\par As the authors point out in \cite{Gao2020China}, the majority of related works in channel modelling for maritime communications mainly rely on large scale channel measurement and analysis, that can be extremely time consuming and costly. Therefore, an alternate approach is presented in that work and evaluated, based on the ray-tracing algorithm, in order to model the channel from a UAV to a moving vessel. In the same context, the surface morphology is simulated and then the ray-tracing algorithm is applied. Simulation results indicate that the amplitude of the received signal follows the Rice distribution. Although results did not take into consideration sea surface motion, experimental evaluation verified the accuracy of the proposed simulation setup. 
\par In \cite{Rasheed2022ieee_trans} the goal is to incorporate 5G mm-Wave with UAV-aided MCNs. To this end, a channel estimation technique based on ML is proposed, that can be used to a wide variety of applications and surrounding environments, via a fully decentralized generative learning model. Simulation results verify the improved accuracy and downlink data rates of the proposed channel estimation framework. Finally, in \cite{Liu2021JCN}, the atmospheric ducting effect is described, that can have a severe impact on the quality of wireless communications especially for long ranges. As a result, UAVs, as flying BSs or mobile terminals, may face remote interference, which can be more severe than the corresponding of the traditional terrestrial nodes.

\subsection{Data rate maximization}

In the vast majority of related works to data rate maximization, the optimization problem is initially formulated and then solved with computationally efficient techniques. In this context, in \cite{Pottoo2022}, the authors investigate a UAV-enabled FSO communication system that could enable a wideband data link with a supporting vessel, when deployed in the Arctic region. To this end, the deployed orientation consists also of a detection unit that can be used for SAR purposes with the help of the UAVs. In the same context, the effects of weather conditions were considered during performance evaluation. According to the presented results, the coexistence of wind and snow had the most severe effect on the FSO link performance followed by fog, snow, and wind, separately.
In \cite{Cao2021}, the authors formulate a mmWave-based UAV-aided MCN optimization problem. In this context, a least mean square (LMS) distributed beamforming algorithm is considered, where transceiver optimization is jointly performed. Simulation results demonstrate the superiority of the proposed normalized approach with respect to the conventional LMS method, both in terms of convergence and spectral efficiency (SE). In the latter case, SE improvement is more evident in the high signal-to-noise ratio (SNR) region. 
\par In \cite{Wang2020}, the problem of on-demand coverage for maritime hybrid satellite-UAV-terrestrial networks is investigated. To this end, a user-centric clustering approach has been considered, where each vessel is served by a specific group of UAVs and terrestrial BSs (TBSs). Moreover, an edge server has been deployed as well in order to perform central control and signal processing. An optimization framework has been proposed with improved convergence, where the minimum user rate served by TBSs and UAVs is maximized, while guaranteeing the leakage interference to satellite users below an acceptable threshold.
In \cite{Ghanbari2022}, a two hop underwater data transfer transmission scheme has been considered, with the help of UAVs, UUVs, and a buoying relay. In this context, the authors analyze the importance of designing an optimal beamwidth in order to achieve the lowest outage probability of the transceiver link. According to the presented analysis, in scenarios with increased transmission power a wider beamwidth should be employed, in order to mitigate the effects of pointing error. On the contrary, when attenuation is the dominant signal deterioration factor, a narrower beam should be employed. 
\par Finally, in \cite{Wang2020ieee_internet}, the same transmission approach as in \cite{Ghanbari2022} has been considered, where analytical path connectivity expressions have been derived, both for the underwater and air link cases. In particular, a UAV-assisted underwater data acquisition scheme is proposed and evaluated in terms of achievable connectivity, where multiple sink nodes deployed on the water surface serve as intermediate relays between under-
water sensors (IoT nodes), deployed either as fixed-grid topology or random, and UAVs. The sink nodes receive acoustic signals that are then forwarded to the UAVs via wireless links. An analytical model is derived, validated by simulation results, that calculates path connectivity, taking into account various vital parameters, such as UAV's trajectory, antenna characteristic, stability of sink nodes, etc. Moreover, a UAV-assisted data acquisition edge computing scheme has been proposed as well.
Simulation results verify the effectiveness of the proposed approach, where overall performance is evaluated for a number of transmission related factors, such as the underwater environment and the antenna beamwidth. 

\subsection{Interference mitigation}

Due to spectrum scarcity, UAVs and MCNs may share the same bandwidth areas. Hence, advanced co-channel interference (CCI) mitigation techniques are required, which is expected to be an unstable factor due to the continuous motion of UAVs. In this context, in \cite{Li2020wc}, various challenges are presented towards the integration of UAVs and MCNs. An important aspect is that vessels follow specific sea lanes and are distributed over large geographical areas. Therefore, on-demand coverage can be provided with the proper placement of UAVs. To this end, the authors analyze various constraining factors, such as the harsh maritime environment which is strongly affected by weather conditions, the difficulty for UAVs to land and charge, as well as the importance of coordination among terrestrial and satellite networks. In the latter case, when UAVs are far away from the coastal area, satellites could be the only choice for wireless backhaul with inevitable large delay and limited communication rate.  Finally, as the authors correctly point out, the trajectory of the UAVs can be exploited in order to extract a CCI pattern. 
In \cite{Fang2020} the authors consider a similar topology as the one in \cite{Wang2020}, where a power allocation strategy is proposed and evaluated based solely on large-scale CSI measurements. To this end, a two-step iterative algorithm has been developed, that converges quickly to the desired solution. The SE can be improved when compared with other power allocation schemes. 
\par Finally, in \cite{Liu2021conf}, a DRL approach is presented and evaluated for jamming mitigation in UAV-aided MCN. In this context, two neural networks are properly trained for power allocation and message transmission. According to the presented results, the proposed approach can significantly improve BER and reduce overall power consumption. 

\subsection{Energy-efficient transmissions}

\begin{figure}[t]
\centering
\includegraphics[width=0.9\columnwidth]{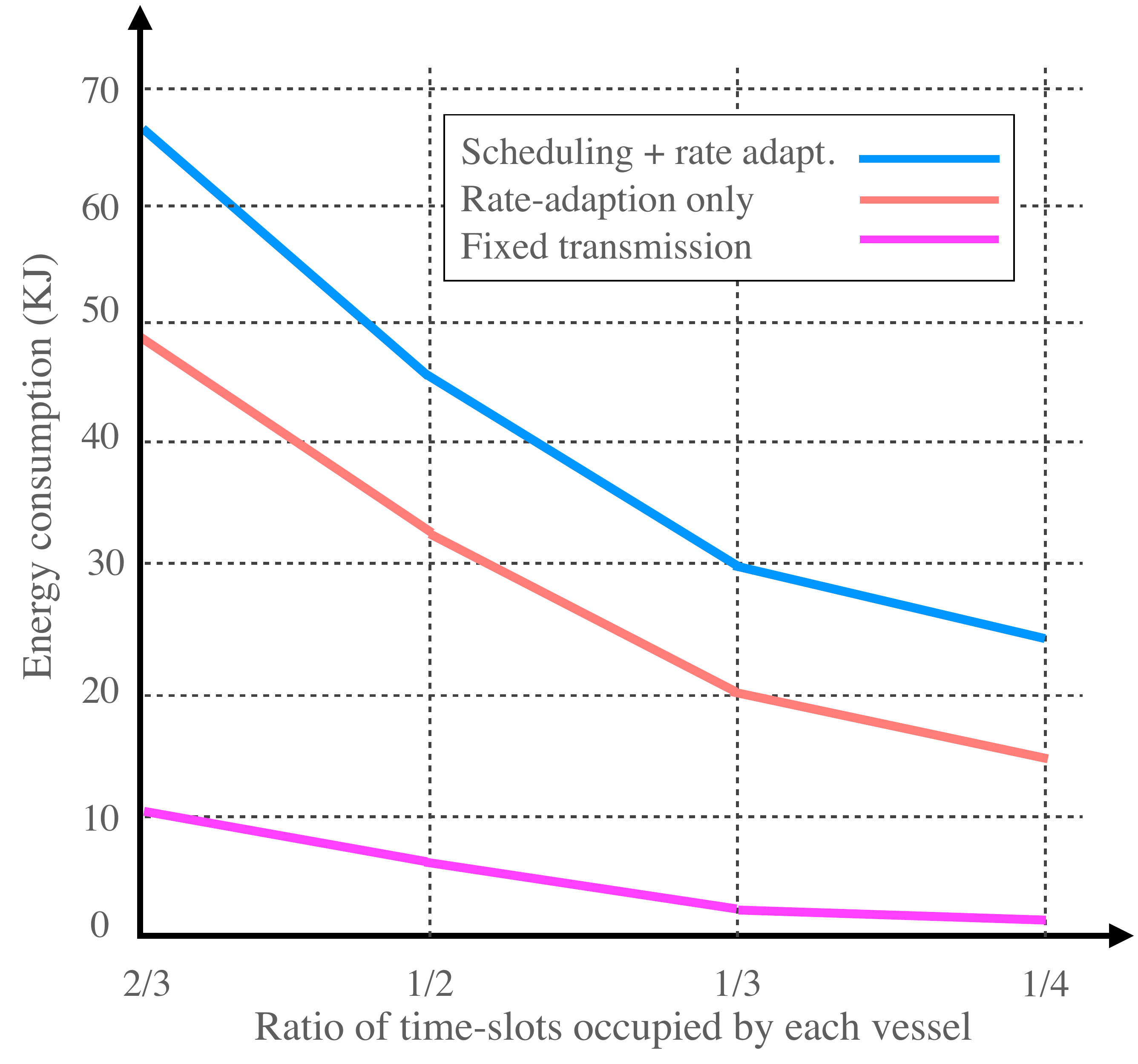}
\caption{Energy consumption of the network for different cases of occupied time-slots for each vessel \cite{Wang2021jsac}.}
\label{wang_jsac}
\end{figure}

The design and deployment of energy efficient communications in UAV-aided MCNs is a challenging research field, since due to the harsh maritime environment the lifetime of individual components, such as UAVs or sensor nodes should be maximized in order to minimize outage probability. In this context, in \cite{Rahimi2022} the authors propose a novel approach for the provision of seamless connectivity in a maritime orientation. To this end, a series of connected buoys is deployed over a wide sea-range in order to handle transmission among UAVs and the on-shore data-fusion and control center. The basic goal is to reduce energy consumption by avoiding unnecessary handover triggers. Therefore, the authors propose a handover decision model based on received signal strength. In this context, the formulated problem, that takes into consideration SNR, available data rate, as well as residual energy and handover data, is solved with the help of a probabilistic-based genetic algorithm. Simulation results indicate that the proposed approach can significantly reduce handover triggers when compared to other approaches.
In \cite{Wang2021jsac}, the authors consider a hierarchical satellite-UAV-terrestrial MCN. To this end, the joint link scheduling and rate adaptation problem for this hybrid network is addressed, in an effort to minimize the total energy consumption with QoS guarantees. A key novelty of the proposed approach is that only the slowly-varying large-scale CSI is considered, which can be obtained easily according to the trajectories of UAVs and the shipping lanes of vessels. Performance evaluation is based on reduced complexity approximations, such as the Min-Max transformations. Results indicate that the proposed approach can converge quickly with moderate computational complexity and has a similar performance compared to the optimal solution. Moreover, as it can be shown in Fig.~\ref{wang_jsac}, significant energy consumption improvement over other approaches is obtained, for different cases of occupied time-slots by each vessel.
Similar to \cite{Wang2021jsac}, the authors in \cite{Li2020ieeetrans} also consider large-scale CSI during the joint optimization of trajectory and in-flight transmit power, subject to constraints on UAV kinematics, tolerable interference, backhaul, and the total energy of the UAV for communications.  
\par Finally, in \cite{Hong2021ieeenetwork}, the authors propose an efficient adaptive antenna design, that can alter the produced radiation pattern according to the supported application. This antenna is composed of cylindrical patch antenna sub-arrays and connected by flexible substrate. In the same context, the authors also highlight the importance of radar-communication integration in the context of MCNs based on 6G technology. 

\section{Radio-Resource Management}\label{RRM}
Proper RRM in UAV-aided maritime communications includes a variety of diverse design goals, such as the maximization of spectral efficiency, proper deployment of UAVs in order to provide efficient network coverage and network lifetime maximization to avoid potential outage especially in long distances from terrestrial networks. Finally, proper authentication mechanisms are of utmost importance as well, since due to the interconnection of a large number of IoT devices, security breaches are more likely to appear. In the following subsections, related works in the aforementioned areas are described, while key considerations are outlined in Table IV. 


\begin{table}[t]
\caption{List of papers on resource management for UAV-aided maritime communications.}
\label{RRM_table}
\centering
{\begin{tabular}[t]{m{7em}<{\raggedright} m{4.4em}<{\raggedright} m{6.9em}<{\raggedright} m{6.7em}<{\raggedright}}
\hline
\rowcolor{whitesmoke}\textbf{Reference}	& \textbf{Maritime topology} & \textbf{Resource management target}  & \textbf{Method}\\
\hline%
\rowcolor{beaublue}\vspace{0.1cm} Ma et al.~\cite{Ma2020network} &\vspace{0.1cm} UAV-aided &\vspace{0.1cm} Network lifetime maximization via relay transmission &\vspace{0.1cm} Optimization based on network lifetime maximization \\%
\rowcolor{whitesmoke}\vspace{0.1cm} Chen et al.~\cite{Chen2020MDPI} &\vspace{0.1cm} UAV-aided &\vspace{0.1cm} Delay and message loss minimization via relay transmission &\vspace{0.1cm} System-level simulations \\%
\rowcolor{beaublue}\vspace{0.1cm} Che et al.~\cite{Che2021conf} &\vspace{0.1cm} UAV-aided &\vspace{0.1cm} Reduction of transmission delay via UAV-assisted relay transmission &\vspace{0.1cm} Iterative optimization\\%
\rowcolor{whitesmoke}\vspace{0.1cm} Kavuri et al.~\cite{Kavuri2020IEEE} &\vspace{0.1cm} UAV-aided &\vspace{0.1cm} Delay and loss probability &\vspace{0.1cm} Simulation framework \\%
\rowcolor{beaublue}\vspace{0.1cm} Lyu et al.~\cite{LyuSpringer} &\vspace{0.1cm} UAV-aided &\vspace{0.1cm} Relay transmission &\vspace{0.1cm} Joint optimization of power and trajectory \\%
\rowcolor{whitesmoke}\vspace{0.1cm} Liu et al.~\cite{Liu2021conf2} &\vspace{0.1cm} UAV-USV-aided &\vspace{0.1cm} Task allocation &\vspace{0.1cm} matching-coalition game \\%
\rowcolor{beaublue}\vspace{0.1cm} Jia et al.~\cite{Jia2021jsac} &\vspace{0.1cm} UAV-USV-aided &\vspace{0.1cm} Heterogeneous networks collaboration &\vspace{0.1cm} matching algorithm \\%
\rowcolor{whitesmoke}\vspace{0.1cm} Cao et al.~\cite{Cao2020conf} &\vspace{0.1cm} UAV-aided &\vspace{0.1cm} Delay minimization, network throughput maximization &\vspace{0.1cm} DRL \\%
\rowcolor{beaublue}\vspace{0.1cm} Jia et al.~\cite{Fang2022trans} &\vspace{0.1cm} Satellite-UAV-Terrestrial &\vspace{0.1cm} Multiple-tier networks cooperation &\vspace{0.1cm} Matching algorithm \\%
\rowcolor{whitesmoke}\vspace{0.1cm}Tang et al.~\cite{Tang2021China} &\vspace{0.1cm} UAV-aided &\vspace{0.1cm} NOMA-aided throughput maximization &\vspace{0.1cm} Problem decomposition \\
\rowcolor{beaublue}\vspace{0.1cm}Ma et al.~\cite{Ma2021internet} &\vspace{0.1cm} UAV-USN-aided &\vspace{0.1cm} NOMA-aided network lifetime maximization &\vspace{0.1cm} weight-based matching \\
\rowcolor{whitesmoke}\vspace{0.1cm} Tang et al.~\cite{Tang2021China} &\vspace{0.1cm} UAV-aided &\vspace{0.1cm} Cooperative data collection &\vspace{0.1cm} Problem decomposition  \\%
\rowcolor{beaublue}\vspace{0.1cm}Xie ~\cite{Xie2018conf} &\vspace{0.1cm} UAV-aided &\vspace{0.1cm} Routing optimization in FANETs &\vspace{0.1cm} ns-3 based evaluation \\
\rowcolor{whitesmoke}\vspace{0.1cm}Chaundry et al.~\cite{Chaudhry2021trans} &\vspace{0.1cm} Satellite-UAV-Terrestrial &\vspace{0.1cm} Security enhancement &\vspace{0.1cm}real or random oracle model \\
\rowcolor{beaublue}\vspace{0.1cm}Khan et al.~\cite{security_elsevier} &\vspace{0.1cm} Satellite-UAV-Terrestrial  &\vspace{0.1cm} Security enhancement &\vspace{0.1cm}Elliptic curve cryptography \\
\hline
\end{tabular}}
\end{table}

\subsection{Cooperative UAV data relaying}
As previously mentioned, UAVs can have a dual role in MCNs: serve either as active BSs in order to provide coverage in distant areas, or act as relay nodes that enhance and re-transmit signals mainly coming from serving nodes located either below or on the surface of the sea. In this context, another typical application of UAV-assisted wireless communications in the maritime sector is for oceanic monitoring networks (OMNs), since they are not restricted by the geographical environment and have advantages on flexible networking configurations. To this end, such an architectural approach, as also mentioned in Section II, consists of two segments, the underwater transmission between underwater sensor nodes (USNs) and sea surface sink nodes (SNs) as well as the RF part, from SNs to ground base stations or UAVs \cite{Ma2020network}. Therefore, UAVs can act as relay nodes among USNs, SNs, and TBSs.   
\par In \cite{Chen2020MDPI}, the performance of a buoy communication mode selection strategy has been
evaluated, where UAVs can act as relay nodes. All UAV-related optimization issues (i.e., trajectory, power consumption, time slot allocation, etc), are separately treated in order to relax computational burden. According to the presented results, the proposed algorithm can significantly improve the minimum throughput of the ocean surface drifting buoys.
In \cite{Che2021conf}, an iterative optimization algorithm has been proposed that reduces average transmission delay and increases transmission success ratio, taking into consideration the motion of container vessels. In \cite{Kavuri2020IEEE} a narrow-band IoT infrastructure is used for tracking containers transported by marine cargo vessels, while operating near the coastline. To this end, UAVs are used as intermediate relay nodes. Extensive system-level simulations were performed indicating that the relay-assisted approach can significantly improve link quality with respect to the standard vessel-BS connection. In \cite{LyuSpringer}, the authors investigate a multi-UAV assisted cooperative transmission to maximize the total throughput under the constraints of outage probability, transmit power and available channels. To this end, a heuristic algorithm is employed to solve the optimization problem. A significant novelty of the presented work is that performance evaluation considers in general a large number of USVs and UAVs compared to other state of the art approaches (i.e., 15 UAVs, 50 USVs). According to the presented results, the performance of the proposed approach is quite close to the optimal solution. In \cite{liu2021uav}, a UAV-USV cooperative communication scenario in smart ocean networks has been investigated with the scope to improve the quality and reduce the cost. To this aim, a matching algorithm has been proposed to  transform the task assignment into a one-to-one matching problem between UAVs and USV coalitions.

\par Finally, in \cite{Jia2021jsac}, a more generic cooperative approach among LEO satellites and HAPs is presented, that can leverage space-air-ground wireless communications. Since the complexity of the corresponding optimization problem is significantly increased when compared to other standard 1-tier networks, a satellite-oriented restricted three-sided matching algorithm is proposed to deal with the matching among users, HAPs, and satellites. In the same context, in \cite{Cao2020conf}, an integrated communication network for terrestrial, sea
and high-altitude platform is proposed and evaluated. To this end, the problem of node accessing is formulated under various constraints and solved with the help of DRL. Simulation results demonstrate that the proposed node access mechanism can effectively improve data transfer rate, average device-to-device (D2D) delay and network throughput.

\subsection{Non-orthogonal multiple access} 

\begin{figure}[t]
\centering
\includegraphics[width=0.9\columnwidth]{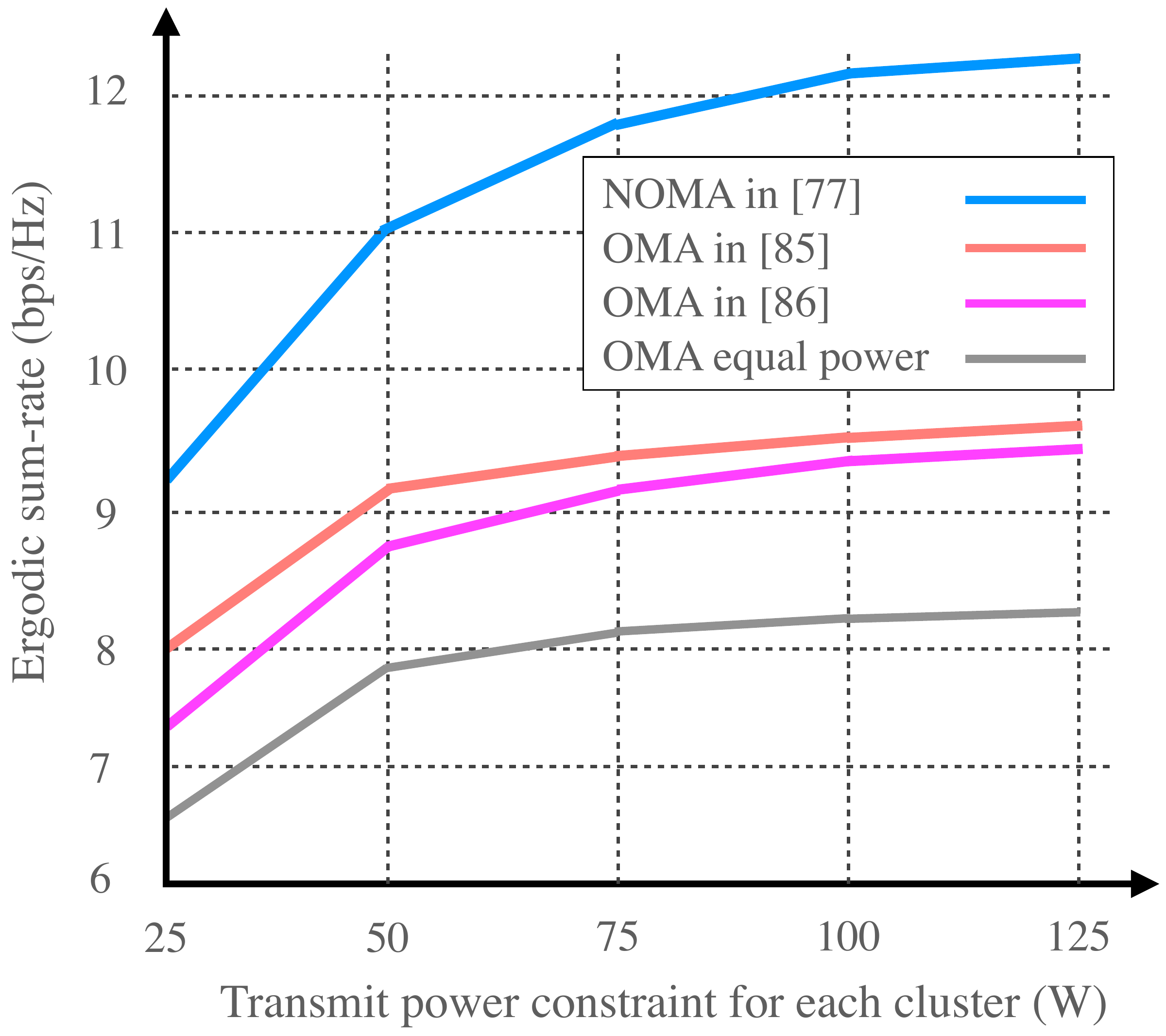}
\caption{Ergodic sum-rate results for various NOMA/OMA power allocation schemes \cite{Fang2022trans}.}
\label{fang_twc}
\end{figure}

In NOMA transmission the same resource block is shared among a group of nodes in order to improve SE and leverage multinode connectivity. In this context, on one hand the NOMA group should be carefully chosen in order to minimize inter-user interference and on the other hand advanced signal reception techniques are required for successive interference cancellation \cite{Nomikos2019ieee_access}. Following a similar approach for hybrid satellite-UAV-terrestrial deployment as the one in \cite{Wang2021jsac} and \cite{Li2020ieeetrans}, where TBSs and UAVs form virtual clusters in a user-centric manner, NOMA transmission has been adopted per cluster in \cite{Fang2022trans}. The spectrum sharing has been considered between nearshore clusters and marine satellites. In this context, a joint power allocation scheme has been proposed, based solely on large-CSI. According to the presented results, the ergodic sum rate has been maximized at a low cost. Fig.~\ref{fang_twc} shows the ergodic sum-rate performance where the superiority of NOMA due to its flexibility in separating users in the power domain over orthogonal multiple access (OMA) alternatives from \cite{feng2013jsac,choi2007twc} is observed. In \cite{Tang2021China}, NOMA groups are formulated by various vessels that are served either by UAVs or TBSs. In order to reduce overall system cost, a single UAV has been considered with a trajectory dictated by the vessels located in the blind zones. In the same context (i.e., cost reduction) both the UAV and vessels are equipped with a single antenna. To solve the non-convex formulated problem, the authors first decompose it into a transmit power allocation subproblem and a transmission duration assignment subproblem. According to the presented results, SE can be improved compared to other approaches.

\subsection{Network lifetime maximization} 

The integration of various technologies in modern MCNs, such as sensing equipment and advanced power supply infrastructure, necessitates a holistic performance evaluation that takes into account the role of various actors in the MCN ecosystem \cite{Carrillo2021ieee_wirel_comm}. Network lifetime maximization is a significant technical challenge that on one hand can reduce outage probability and on the other hand ensure the provision of end to end high data rate applications. To this end, the works in \cite{Ma2021internet} and \cite{Ma2020network} propose an architectural approach where UAVs are used as relay nodes between SNs and TBSs. In this context, and in an effort to improve network's lifetime, resource allocation and UAV deployment issues are modelled as a non-convex optimization problem. This problem is then decoupled to two separate sub-problems: delay minimization for the UAV-SN part and and lifetime maximization for USN-SN communication. Simulation results verified the improved performance of the proposed approach compared to other time division multiple access (TDMA) schemes. 
\par In \cite{Xie2018conf}, a routing protocol based on optimized link-state routing protocol is proposed and evaluated, based on the node link expiration time and residual energy. This approach can significantly improve end-to-end delay, packet transmission rate and routing overhead.

\subsection{Security considerations} 
The integration of various and diverse communication networks in order to form a MCN, has raised security and privacy concerns, especially in the forthcoming 6G-IoT era. In \cite{Chaudhry2021trans}, the authors propose a lightweight authentication protocol for a 6G-IoT MCN. The proposed approach has been evaluated with the help of formal security assessment methods and compared to other security mechanisms. The presented results indicate that the proposed approach achieves a better security-to-efficiency trade-off compared to other state-of-the-art approaches.
In \cite{security_elsevier}, the authors propose a resource friendly authentication scheme for MCNs based on elliptic curve cryptography. According to experimental evaluation, the proposed approach has improved performance compared to other approaches in terms of computational and communication cost. 

\begin{table}[t]
\caption{List of papers on cloud/edge solutions for UAV-aided maritime communications.}
\label{Edge_table}
\centering
{\begin{tabular}[t]{m{7em}<{\raggedright} m{4.4em}<{\raggedright} m{6.9em}<{\raggedright} m{6.7em}<{\raggedright}}
\hline
\rowcolor{whitesmoke}\textbf{Reference}	& \textbf{Maritime topology} & \textbf{Cloud/edge target}  & \textbf{Method}\\
\hline%
\rowcolor{beaublue}\vspace{0.1cm}Dai et al.~\cite{dai2022uav} &\vspace{0.1cm} UAV-USV based &\vspace{0.1cm} Offloading delay &\vspace{0.1cm} Penalty convex-concave
procedure \\
\rowcolor{whitesmoke}\vspace{0.1cm} Hassan et al.~\cite{hassan2021demand} &\vspace{0.1cm} UAV-based &\vspace{0.1cm} Maximize network profit &\vspace{0.1cm} Bender de-
composition \\%
\rowcolor{beaublue}\vspace{0.1cm} Yang et al.~\cite{yang2022multi} &\vspace{0.1cm} UAV-based &\vspace{0.1cm} Time delay and energy consumption &\vspace{0.1cm} Upper confidence bound
\\%
\rowcolor{whitesmoke}\vspace{0.1cm} Zeng et al.~\cite{zeng2020mobile} &\vspace{0.1cm} UAV-based &\vspace{0.1cm} Latency and energy consumption &\vspace{0.1cm} Multi-agent DRL \\%
\rowcolor{beaublue}\vspace{0.1cm} Xu et al.~\cite{xu2020deep} &\vspace{0.1cm} Satellite-
UAV-Terrestrial &\vspace{0.1cm} Resources utilization and latency &\vspace{0.1cm} DRL \\%
\rowcolor{whitesmoke}\vspace{0.1cm} Liu et al.~\cite{9678008} &\vspace{0.1cm} UAV-based &\vspace{0.1cm} Minimization of the latency &\vspace{0.1cm} DRL \\%
\rowcolor{beaublue}\vspace{0.1cm} Zeng et al.~\cite{zeng2022collaborative} &\vspace{0.1cm} UAV-USV based &\vspace{0.1cm} Reduction of the execution time  &\vspace{0.1cm} Incentive-based collaborative \\%
\rowcolor{whitesmoke}\vspace{0.1cm} Hassan et al.~\cite{hassan2021blue} &\vspace{0.1cm} Satellite-
UAV-Terrestrial &\vspace{0.1cm} Weighted computational/ communication sum-rate  &\vspace{0.1cm} Bender and primal decomposition \\%
\rowcolor{beaublue}\vspace{0.1cm} Dai et al.~\cite{dai2022deep} &\vspace{0.1cm} UAV-based &\vspace{0.1cm} Minimization of the mission time &\vspace{0.1cm} DRL\\
\hline
\end{tabular}}
\end{table}

\section{Cloud/Edge Computing and Data Offloading}\label{cloud_edge}
The recent mobile communication architecture advances have included MEC as an efficient approach for providing powerful computing capability and low latency to the users. This approach has been also adopted in MCN, where the edge servers, which in several cases are mounted at the UAVs, are deployed close to the ship terminals (or maritime IoT) and have been used to offload computing tasks. Next, related works on mobile edge computing and data offloading in maritime UAV-assisted communication scenarios are discussed, while key considerations are outlined in Table~\ref{Edge_table}.

\begin{figure}[t]
\centering
\includegraphics[width=\columnwidth]{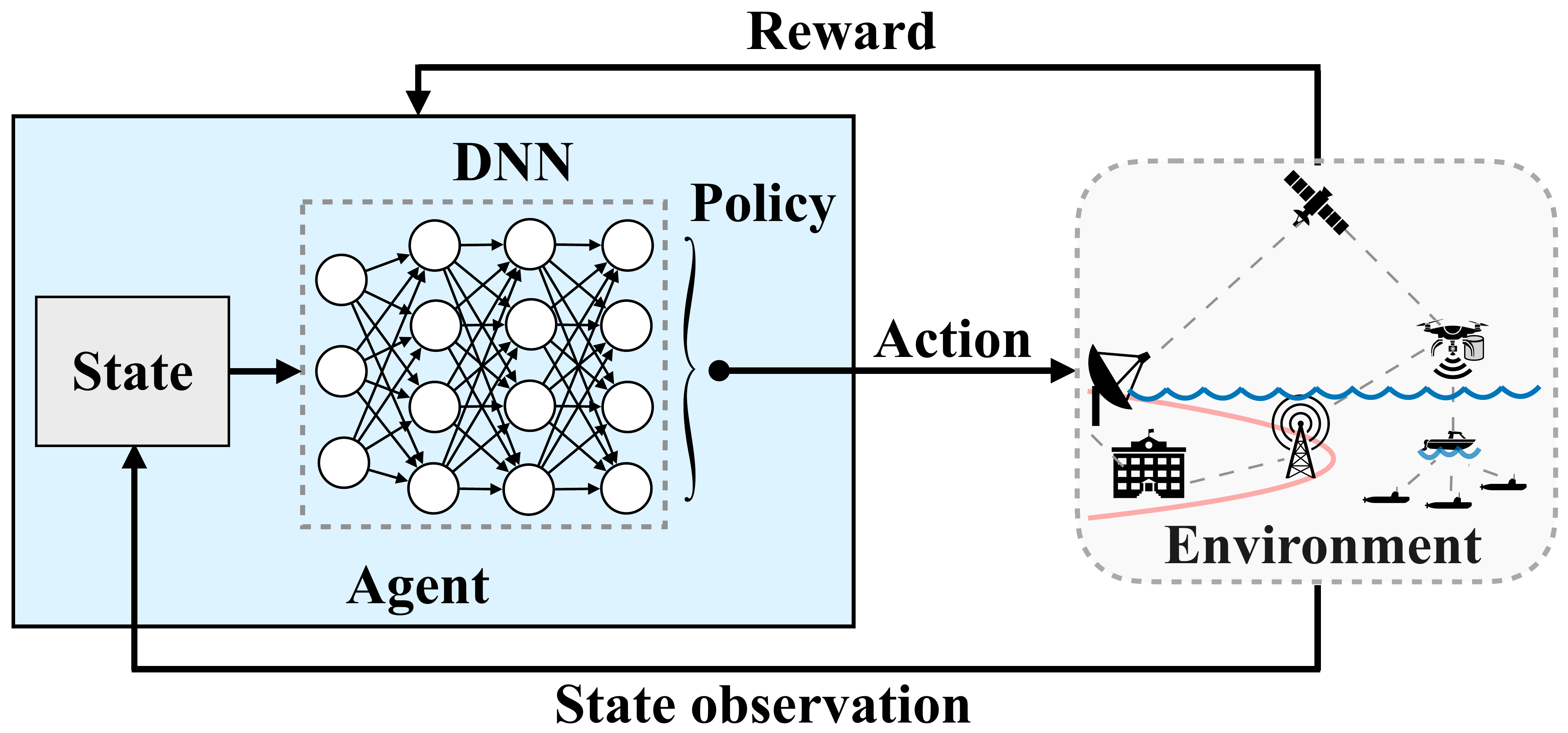}
\caption{DRL architecture in the context of a UAV-aided MCN.}
\label{DRL}
\end{figure}

\subsection{Mobile edge computing}
An important research objective in UAV-assisted MEC studies that is also adopted in MCN is to simultaneously efficient utilize communication, energy resources, by taking also into account latency and computational complexity constraints. These optimization formulations have been also followed in the following papers, which have been solved with the aid of conventional and DRL approaches. In \cite{hassan2021demand}, UAV-assisted maritime IoT communication network has been studied, with the scope to maximize the network profit in MEC-UAVs deployments. The combinatorial optimization problem that has been formulated has been solved using an iterative algorithm that is based on Bender decomposition approach. The simulation results presented, demonstrate the near optimal solutions offered by the proposed approach, in an efficient convergence time. In \cite{zeng2020mobile}, UAV-assisted maritime communication system is investigated, in which UAVs have been employed as mobile data centers in a mobile edge communications, computing, and caching framework. In order to improve the caching rate a multi-agent DRL technology has been used, which predicts users’ preferences and reduces task latency. It is shown that a considerable reduction on the task latency and energy consumption is obtained. In \cite{gao2020multi} and \cite{yang2022multi}, a space-air-ground-edge integrated maritime communications network has been assumed, in which the UAVs can be used as edge servers in order to offload tasks of users. The scope of this paper is to propose an algorithm for selecting the best UAV edge server, which results to a multi-armed bandit problem, taking also into account budget constraints as well as other weighted factors such as delay and energy consumption. Based on the upper confidence bound algorithm, the UAV selection problem can be solved and the resulting simulation results prove the effectiveness of the proposed approach in terms of lower offloading latency and weighted latency-energy
cost. In \cite{xu2020deep}, in a similar system model as it is the one investigated in \cite{yang2022multi}, a joint communication and computation resource allocation problem has been introduced in order to minimize the resource utilization and computational delays. Using a DRL approach for resolving the optimization problem, it resulted to an important improvement of the utilization efficiency of resources as well as a reduction to the reduction of the task implementation latency. Fig.~\ref{DRL} illustrates a centralized DRL architecture where a UAV-aided MCN is optimized following a reward-based procedure, taking appropriate actions to maximize its utility function.

In \cite{9678008}, a two-layer UAV-enabled MEC maritime communication network has been investigated, in which parallel computation of tasks offloading with different amount of data in different virtual machines has been studied. In this framework, a latency minimization problem has been formulated, which takes into account both communication and computation aspects. This problem has been solved using deep Q-network and deep deterministic policy gradient algorithms that aim to find optimal flight trajectories and number of virtual machines participation. The various results presented proved the reduction to the total average latency as compared to conventional approaches. In \cite{hassan2021blue}, a space-air-sea non-terrestrial
network has been studied, in which LEO-MEC satellite allows users to offload data and UAV-MEC is employed for backhauling purposes. In an effort to maximize the weighted computational and communication sum rate, a joint Bender and primal decomposition algorithm has been proposed. Based on this approach an near optimal performance is achieved with an efficient convergence time. In \cite{pang2020space}, a space-air-ground-aqua integrated network architecture is studied, which is combined with MEC technology. In that paper several research challenges were discussed, while various future direction were also proposed.

\subsection{Data Offloading}
In the following data offloading works, the main objective is also a time-related reduction.  
In \cite{dai2022uav}, a UAV-assisted data offloading model is proposed for maritime communications, which takes into account the mobility of container vessel, the UAV movements, and the wireless propagation over the sea. The problem is solved using the penalty convex-concave procedure, and at the various simulation results presented, it was shown that the proposed algorithm can efficiently reduce
the average offloading delay and offloading success
ratio. In \cite{zeng2022collaborative}, a UAV-USV-based collaborative network with maritime cloud servers is introduced. In this collaborative computation offloading framework, the UAVs offloading ratio, to the USVs fleets, is investigated using an incentive-based collaborative
computation offloading scheme. The simulation results indicated a significantly reduction of the computation tasks overall execution time. In \cite{dai2022deep}, a UAV-assisted data collection and data offloading system in marine IoT has been investigated. The objective is to minimize the total mission completion time, by taking into account UAV trajectory, data offloading capabilities, buoy-UAV association relationship, and the transmit power. The mixed integer non-convex problem that is introduced is solved based on a delayed deep deterministic algorithm. The simulation results presented, prove the effectiveness of the proposed scheme for reducing the total mission completion time.


\section{UAV Trajectory Optimization}\label{trajectory}

Trajectory optimization is another important aspect of UAV-aided MCNs, since it is directly related to a considerable number of performance metrics described in the previous sections, such as network lifetime maximization and energy consumption minimization. Moreover, in emergency situations, proper trajectory design can reduce overall network installation time and leverage high quality connectivity. In the following subsections, related works are categorized according to communication performance enhancement and search and rescue procedures. 

\subsection{Enhancing communication performance}

In this subsection various recent works will be analyzed that mainly employ efficient trajectory optimization algorithms towards improving link connectivity under various constraints, such as transmission power and provision of minimum data rate.  In \cite{Velascodronesmdpi} a simulation platform has been developed with the help of Matlab/Simulink for UAV and USV cooperation and interaction with the environment. Therefore, UAV trajectory can be visualised with the help of a flexible software tool that can be executed in various platforms.
In \cite{Zhang2021conf}, a multi-UAV maritime IoT topology has been considered, used to collect information from SNs. In this context, the goal is to find the optimum placement of UAVs and SNs that minimizes overall energy consumption. The integer linear programming optimization problem is formulated and then solved with the help of the Gurobi solver. In \cite{Guan2021conf}, the optimum placement of UAVs is investigated for off-shore relay communications. To this end, a minimum-maximization optimization problem of link capacity is formulated and solved with the help of particle swarm optimization (PSO) algorithm. \par 
In \cite{Zhang2020confglobecom} - \cite{Zhang2022trans}, a non-convex optimization problem is formulated in order to reduce the energy consumption in a UAV when used as a relay node that gathers and processes information from multiple deployed buoys. In this context, the communication time scheduling among the buoys and the UAV's flight trajectory subject to wind effect are jointly optimized. The formulated optimization problem is solved with the help of the successive convex approximation (SCA) method, based on a proposed cyclical trajectory design framework that can handle arbitrary data volume efficiently subject to wind effect.
\par In \cite{Liu2022} a MCN is considered with UAVs, buoys, and underwater sensors. The goal is to maximize energy efficiency (EE) by jointly optimizing the transmit power of buoys and sensors, scheduling their transmissions, as well as designing the UAV's trajectory. To this end, three separate sub-problems are considered in an effort to relax computational burden, that are solved with the help of SCA method. According to the presented results, the adopted method has improved convergence time when compared with other methods.
In \cite{Lyu2022conf}, the authors evaluate the performance of fast UAV trajectory planning algorithm based on the Fermat-point theory for a maritime IoT with USVs. Results indicate that the proposed approach can significantly improve data collection rate of USVs and provide fast convergence at the same time. In \cite{Li2021conf}, a multi-relay orientation has been considered, where the goal is trajectory optimization with the help of dual Q-learning. In this context, the objective is to minimize the total average loss and thus improve link quality. In \cite{Li2019conference}, considering only large-scale CSI, an algorithm is proposed based on problem decomposition and successive convex optimization for effective power allocation and trajectory planning in a hybrid satellite- UAV-terrestrial MCN. In \cite{Zhang2020chinacommunications}, a UAV relay orientation has been considered, where the goal is to maximize overall data rate. For the special case of a single user, the authors derive a semi closed-form expression of UAV placement. In the same context, an algorithm to find the optimal UAV placement for the general case with multiple users has been presented and evaluated. In \cite{Ho2021conf}, a UAV is used for data collection from various sensor nodes. The goal is to optimize UAV trajectory in order to minimize total energy consumption, maximize data rate along with the flight duration of the UAV. The optimization problem is formulated and solved with the help of PSO, for various CSI scenarios (i.e., full or limited sensor node position information). To this end, a Kalman filter is used to improve the position estimation errors. Finally, a similar approach with PSO modelling has been also followed in \cite{Yan2021} to solve the random task allocation problem of multiple UAVs and the two-dimensional route planning of a single UAV.

\subsection{Improving maritime search and rescue}

UAVs can be alternately deployed for search and rescue purposes in MCNs, where the primary goal is to reduce overall network delay.  
In \cite{Kim2018}, various research challenges are presented and investigated in the field of wide are monitoring in MCNs with the help of UAVs, such as optimum number and deployment of UAVs as well as scheduling algorithms for patrolling and recharging UAVs. 
In \cite{Liu2021conf2} the authors consider a maritime emergency communication scenario, where a UAV acts as a BS and can communicate with a number of active nodes in the sea. The goal is to jointly optimize power allocation and trajectory of the UAV, in order to maximize system's throughput. In such an emergency scenario, the received signal from an arbitrary node is broadcasted to the individual users within the close proximity of the node. The optimization problem is formulated and solved with the help of a loop iterative algorithm. In \cite{ZuoBeidou2020}, an extended search algorithm is presented and evaluated with the help of MATLAB for efficient and rapid maritime search. The presented results reveal a close relationship among the height of the UAV and the overall rescue time. 
In \cite{Oliveira2019conf}, a computational tool has been presented and evaluated towards leveraging communications range and capacity limits of ad-hoc networks of UAVs operating in maritime scenarios. To this end, performance evaluation consisted of two scenarios: collaborative search and tracking of targets and circular formation flight for detection of external threats to ship convoys.
In \cite{Yang2020}, a cooperative communication model among UAVs and USVs in MCNs is analyzed, for SAR purposes. In this context, RL is used to plan the optimum search path and improve overall throughput. The authors evaluated achievable throughput for various reward functions in order to improve the data throughput of the system. 
\par Moving a step forward, in \cite{Wu2020}, the problem of underwater object detection is investigated, via the collaboration of UAVs, USVs, and UUVs. In this context, the overall strategy is divided into the search phase and the track phase. In each one of the two phases, the goal is to maximize the search space and minimize the terminal error respectively. An improved PSO algorithm has been deployed, that can be executed either in a centralized or in a distributed mode. According to the presented results, the joint UAV-USV-UUV integrated system can be more efficient from the USV-UUV for the search and track procedures.
In \cite{Brown2020}, the authors optimize the trajectory of a UAV when deployed for maritime radar wide area persistent surveillance with the objective goals to minimize power consumption maximize mean probability of detection, and minimize mean revisit time. To this end, a multiobjective PSO algorithm has been employed and evaluated for two realistic operational scenarios.


\begin{table}[!t]
\caption{List of papers on UAV trajectory design for UAV-aided maritime communications.}
\label{Trajectory_table}
\centering
{\begin{tabular}[t]{m{7em}<{\raggedright} m{4.4em}<{\raggedright} m{6.9em}<{\raggedright} m{6.7em}<{\raggedright}}
\hline
\textbf{Reference}	& \textbf{Maritime topology} & \textbf{Trajectory optimization target}  & \textbf{Method}\\
\hline%

\rowcolor{beaublue}\vspace{0.1cm} Velasco et al.~\cite{Velascodronesmdpi} &\vspace{0.1cm} UAV-based &\vspace{0.1cm} Reduction of development and delivery times &\vspace{0.1cm} Simulink \\%

\vspace{0.1cm} Zhang et al.~\cite{Zhang2021conf} &\vspace{0.1cm} Multi-UAV-based &\vspace{0.1cm} Energy consumption &\vspace{0.1cm} Gurobi solver \\%

\rowcolor{beaublue}\vspace{0.1cm}Guan et al.~\cite{Guan2021conf} &\vspace{0.1cm} UAV-based &\vspace{0.1cm} Link capacity &\vspace{0.1cm} PSO \\

\vspace{0.1cm}Zhang et al.~\cite{Zhang2020confglobecom}, \cite{Zhang2022trans}&\vspace{0.1cm} UAV-based &\vspace{0.1cm} Energy consumption &\vspace{0.1cm} SCA \\

\rowcolor{beaublue}\vspace{0.1cm}Zhixin et al.~\cite{Liu2022}&\vspace{0.1cm} UAV-based &\vspace{0.1cm} Joint resource allocation and trajectory design &\vspace{0.1cm} SCA \\

\vspace{0.1cm}Lyu et al.~\cite{Lyu2022conf}&\vspace{0.1cm} UAV-USVbased &\vspace{0.1cm} Data collection rate of USVs&\vspace{0.1cm} Fermat-point theory \\

\rowcolor{beaublue}\vspace{0.1cm}Li et al.~\cite{Li2021conf}&\vspace{0.1cm} UAV-based &\vspace{0.1cm} Losses minimization &\vspace{0.1cm} Dual Q-learning \\

\vspace{0.1cm}Li et al.~\cite{Li2019conference}&\vspace{0.1cm} Hybrid Satellite-UAV-Terrestrial &\vspace{0.1cm} Maximization of minimum data rate &\vspace{0.1cm} Successive convex optimization \\

\rowcolor{beaublue}\vspace{0.1cm}Zhang et al.~\cite{Zhang2020chinacommunications}&\vspace{0.1cm} UAV-assisted &\vspace{0.1cm} Data rate maximization &\vspace{0.1cm} One-dimensional linear search \\

\vspace{0.1cm}Ho et al.~\cite{Ho2021conf}&\vspace{0.1cm} UAV-assisted &\vspace{0.1cm} Energy minimization, throughput maximization &\vspace{0.1cm} Kalman Filter, PSO \\

\rowcolor{beaublue}\vspace{0.1cm}Yan et al.~\cite{Yan2021}&\vspace{0.1cm} UAV-assisted &\vspace{0.1cm} Task allocation &\vspace{0.1cm} PSO \\

\vspace{0.1cm}Liu et al.~\cite{Liu2021conf2}&\vspace{0.1cm} UAV-based &\vspace{0.1cm} Throughput maximization &\vspace{0.1cm} Loop iterative algorithm \\

\rowcolor{beaublue}\vspace{0.1cm}Zuo et al.~\cite{ZuoBeidou2020}&\vspace{0.1cm} UAV-based &\vspace{0.1cm} Search time improvement &\vspace{0.1cm} Square search/MATLAB \\

\vspace{0.1cm}Oliveira et al.~\cite{Oliveira2019conf}&\vspace{0.1cm} UAV-based &\vspace{0.1cm} collaborative search and tracking &\vspace{0.1cm} Simulation tool \\

\rowcolor{beaublue}\vspace{0.1cm}Yang et al.~\cite{Yang2020}&\vspace{0.1cm} UAV-USV-based &\vspace{0.1cm} collaborative search and rescue &\vspace{0.1cm} RL \\

\vspace{0.1cm}Wu et al.~\cite{Wu2020}&\vspace{0.1cm} UAV-USV-UUV based &\vspace{0.1cm} collaborative SAR &\vspace{0.1cm} PSO \\

\rowcolor{beaublue}\vspace{0.1cm}Brown et al.~\cite{Brown2020}&\vspace{0.1cm} UAV-based &\vspace{0.1cm} Trajectory optimization &\vspace{0.1cm} PSO \\

\hline
\end{tabular}}
\end{table}

\section{Experimental Studies}\label{experimental}

In this section representative recent works are analyzed that deal with experimental evaluation of UAV-aided MCNs, considering various design aspects as described in the previous sections. In this context, in \cite{Sollesnes2018conf} preliminary results are presented towards the development of an autonomous ocean observing system using miniature underwater gliders that can operate with the support of UAVs and USVs for deployment, recovery, battery charging, and communication relay. According to the presented results, the individual components are shown to be pressure tolerant retaining functionality at pressure equivalent to 200m depth. In \cite{Yokota2021mdpi}, experimental demonstrations have been performed in order to evaluate the feasibility of using UAVs as a sea-surface base for underwater communication with an UUV. According to the presented results, UAVs can provide an efficient communication link for distances near the shore, achieving at the same time robust hovering control.

In \cite{Yu2022conf}, a channel measurement campaign was performed for the communication between a UAV and an USV at the S-band. The analysis included large-scale and small-scale channel characteristics, including path-loss, shadow fading, and multipath fading.
In \cite{Pokorny2021sensors}, an experimental study is presented concerning the deployment of a USV and a UAV in an autonomous collaborative communication system. To this end, two communication scenarios are evaluated, the first one including a direct link among the TBS and the USV, while the second one considers the UAV as a relay node with respect to the aforementioned link. To this end, directional antennas are placed on the UAV and the USV with the appropriate steering mechanisms to align radiation patterns. 
In \cite{Teixeira2020ieeeaccess}, the authors evaluate the RF propagation at MCNs using the BLUECOM+ solution, which consists of a multi-hop aerial backhaul network. To this end, a height control approach has been presented and evaluated, taking into account all reflected signals in order to maximize reception quality. The relay node positioning problem was defined as an optimization problem and solved using the PSO technique. According to the presented results the PSO approach outperformed the trivial methods in terms of overall throughput, such as fixed or random height selection. 
\par In \cite{Guldenring2019} a UAV-aided MCN network is evaluated when used for rescue purposes, where applicability of Long Term Evolution (LTE) is investigated. At the first step, a detailed maritime channel model is developed and implemented. The model is then evaluated via hardware implementation. In \cite{Gorczak2019}, potential benefits from the integration of UAVs in SAR operations with LTE networks are evaluated. To this end, the authors develop a resource-guaranteed scheme based on persistent scheduling, using an open-source LTE stack. The approach is evaluated with a laboratory setup using software-defined radio modules. According to the presented results, latency and reliability can be significantly improved. In \cite{drones4020016}, a novel multi-link approach based on LTE networks is proposed in an effort to increase overall network throughput by aggregation. Moreover, large scale experiments have been conducted and published. According to the presented results, the adopted multi-link approach can enable smooth handovers between different networks. In \cite{Ribeiro2018}, field trials were performed concerning the cooperation between a UAV and a USV. Finally, in a similar context in \cite{Zolich2017}, an experimental evaluation took place considering various UAVs, USVs, and UUVs. To this end, various operational scenarios were considered. In all scenarios, including direct and relayed connection, the network was found to be reliable.

\section{Open Issues}\label{open_issues}

The integration of UAVs in MCNs can provide tremendous gains in coverage, delay reduction, reliability and deployment flexibility. As research on this areas has only recently started, there are several open issues and interesting research directions to explore.

\subsection{Physical-layer issues}

\par The deployment of a large number of transmitting antennas results in massive MIMO (mMIMO) configurations. In general, such configurations can provide improved spatial separation among active nodes via the generation of highly directional lobes. Therefore, both SE and EE can be leveraged \cite{Elhoushy2022}. However, in a MCN orientation, the deployment of mMIMO antennas in UAVs would result in increased installations costs and overall hardware complexity. In this context, decentralized architectures (d-MIMO) are an active area of research interest, as significant advantages compared to the centralized structures can be provided \cite{iccdmimo}. In particular, d-MIMO systems can reduce spatial correlation, increase the diversity and multiplexing gain, and reduce the average path-losses through effectively shortening transmission distances, which can be beneficial in MCNs. Another promising technique is related to IRS-aided UAV networks. In such cases, wireless channel quality can be significantly enhanced through the deployment of UAVs carrying IRSs that appropriately perform phase shifting of the relayed signals \cite{nguyen2022tccn, shang2021wcm}. 
\par Finally, an interesting research topic is the design and implementation of a measurement campaign for 6G carrier frequencies. So far, even up to date works, such as the one presented in \cite{Liu2021JSAC} have considered frequencies up to 5GHz during performance evaluation.

\subsection{UAV-aided non-orthogonal multiple access}

UAV-aided MCNs possess high flexibility due to the existence of LoS conditions and the re-positioning capabilities of UAVs. As NOMA schemes have been increasingly popular in recent years, due to their potential to improve the performance of mobile networks, the development of NOMA solutions for UAV-aided MCNs is an important research area. Such algorithms should optimize the UAV trajectory and positioning in order to maintain channel asymmetry between co-existing network nodes, thus maximizing the spectral efficiency of the transmission \cite{liu2022jsac}. Moreover, the integration of NOMA in mobile edge networks, where task allocation, caching location and power allocation for NOMA are jointly determined represents another highly efficient solution \cite{ding2018tcom,pei2020tvt}.
Meanwhile, the gains of NOMA in buffer-aided networks have already been highlighted in several studies, further strengthening communication reliability at the cost of a slight delay increase \cite{nomikos2021wc, li2020tcom}.

\subsection{Machine learning}
\begin{figure}[t]
\centering
\includegraphics[width=\columnwidth]{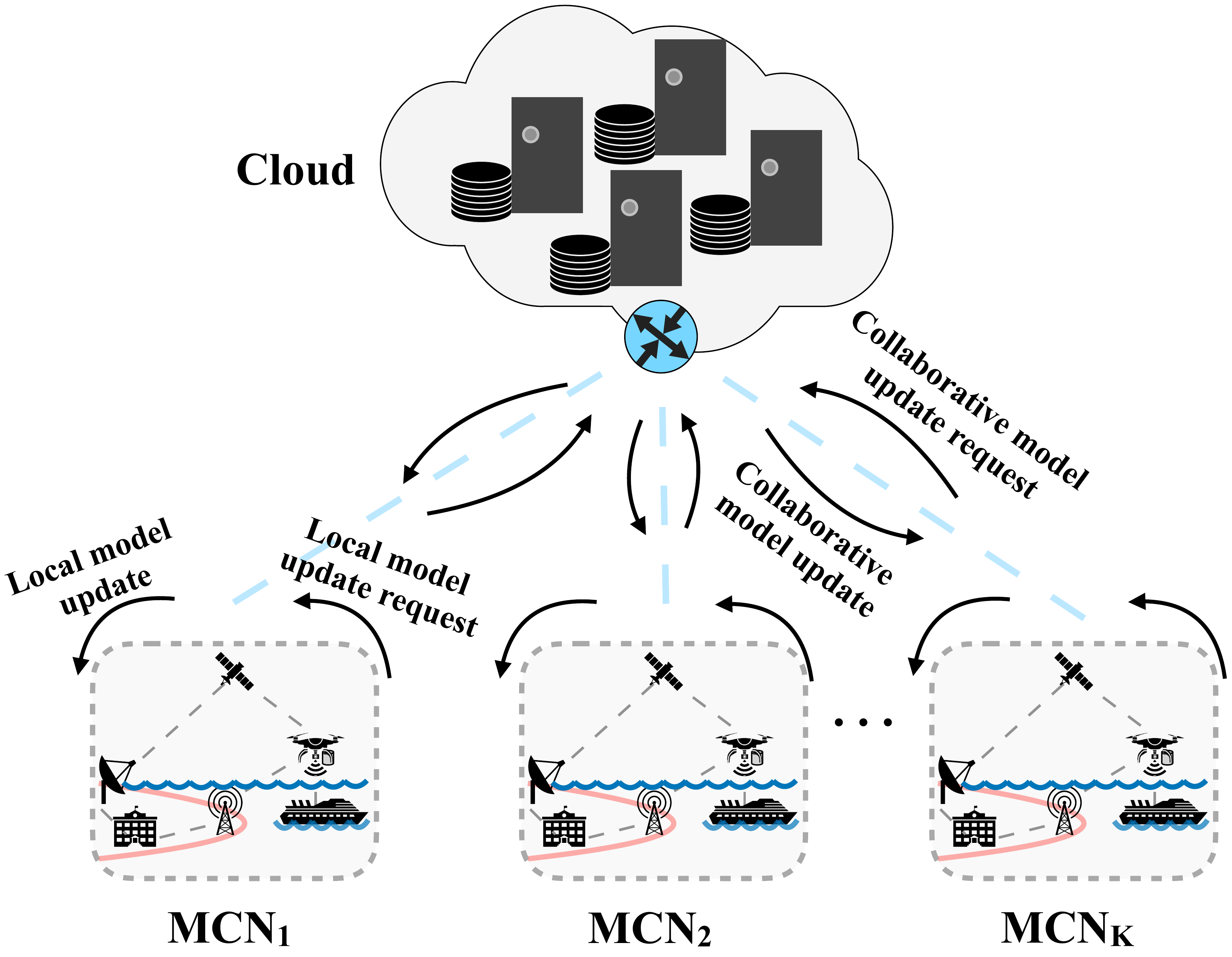}
\caption{Federated learning architecture of a maritime network, comprising $K$ UAV-aided coverage areas.}
\label{FL}
\end{figure}

In the vast majority of related works to key performance indicator optimization in UAV-aided MCNs and performance evaluation, a limited network topology has been considered (i.e., reduced number of UAVs, service nodes, etc) in order to reduce the computational complexity of the employed optimization approaches. However, a more accurate performance evaluation requires large scale orientations to be considered employing a realistic number of active IoT devices. To this end, advanced ML algorithms such as DRL can be quite effective towards process optimization, due to their inherent ability to adapt to various network conditions and perform corresponding adjustments \cite{lei2021network, lahmeri2021ojcoms}. However, since MCNs are spread over a wide territory, edge-based solutions in conjunction with federated learning are a promising solution towards the integration of MCNs in 6G orientations \cite{qu2021network}.  Fig.~\ref{FL} depicts a federated learning architecture where a maritime network comprising $K$ UAV-aided coverage areas, exploits local model updates to improve the efficiency of the global collaborative model.

\subsection{Security and safety}

It is critical to develop secure UAV-aided communication systems to overcome different attack types on communication reliability \cite{challita2019wcm} and the safety of critical infrastructures. In this context, for MCNs relying on UAVs, disrupting the UAV's transmissions by identity forging might be difficult to identify especially in networks of UAV swarms. Thus, UAV authentication and analysis of the received information content is needed, with possible implications for the delay performance \cite{Chaudhry2021trans}. Moreover, in intelligent maritime transportation systems, UAV swarms coordinate to perform various procedures. Still, there exist several vulnerabilities that attackers can exploit towards injecting false data and disputing UAV manoeuvring and collision avoidance systems. In cases where data privacy issues arise, resorting to federated learning is preferable as the federated averaging algorithm builds a global model by aggregating the weighted average of locally updated model at each network device \cite{yang2020twc}. Finally, even though UAV-aided MCNs ripe the benefits of high mobility, low cost, on-demand resource provisioning, and LoS connectivity, they might be vulnerable against eavesdroppers \cite{sun2019wcm, wang2019wcm}. As a result, distributed and low-complexity physical layer security (PLS) algorithms must be developed that will be suited to UAV-aided MCNs, exploiting advancements in the field of ML for autonomous operation \cite{zhang2020tvt}

\subsection{Advanced edge maritime services}

An intrinsic characteristic of current MCNs is the intermittent connectivity due to limited coverage. As maritime services will target at autonomous operation of ships, equipment and unmanned vessels, real-time data processing is of utmost importance. UAV-aided MEC can alleviate this issue, as data will be processed locally, thus avoiding the excessive use of backhaul link \cite{nomikos2022access}. The deep integration of UAVs, USVs, and UUVs can lead to optimal computing task allocation and real-time data collection, thus minimizing network latency. Efficient edge computing and caching algorithms should consider processing capabilities jointly with the trajectories of the unmanned vehicles, and their energy constraints in order to maintain high UAV availability in the network \cite{liu2022ojcoms, tran2022ojvt}.

\subsection{UAV swarm intelligence}

The autonomous operation of UAV-aided MCNs is critical either for providing wireless coverage to ships and unmanned vessels or for edge computing and caching purposes. The ad-hoc formation of UAV swarms requires reliable inter-UAV connectivity and distributed intelligence with UAVs exchanging data regarding their velocity and trajectory in order to avoid collisions and maximize the coverage and task offloading capabilities \cite{wu2022arxiv}. Still, wireless communication issues might arise related to delay in data exchange, channel quality degradation, energy constraints and harsh meteorological conditions affecting UAV stability. In this context, ML-based algorithms e.g. through bandit-based channel prediction can alleviate issues related outdated channel state acquisition \cite{nomikos2022ojcoms} while WPT solutions for UAV swarms can mitigate the impact of energy limitations.

\section{Conclusions}\label{conclusions}

The introduction of UAVs in wireless networks has allowed increased deployment flexibility and dynamic resource provisioning, aligning with 6G ubiquitous connectivity goals. In maritime communications, integrating UAVs to complement shore- and satellite-based deployments provides an intermediate aerial layer that overcomes the limited coverage of terrestrial base stations and increased latency and narrow-band links of satellites. This survey has presented the role of UAVs in maritime network architectures, where they enable a plethora of maritime IoT and broadband service use cases. In addition, various algorithms relying on conventional optimization and machine-learning techniques have been discussed, addressing physical-layer, resource management and cloud/edge computing and caching issues. Furthermore, the important area of UAV trajectory design for maritime communications and search and rescue has been reviewed while current efforts in experimental implementation have been presented. Finally, several important open issues towards efficiently integrating and exploiting UAVs for maritime activities in the 6G era have been outlined, aiming to attract high research interest in this important domain.

\let\oldaddcontentsline\addcontentsline
\renewcommand{\addcontentsline}[3]{}
\bibliography{References}

\begin{thebibliography}{100}
\providecommand{\url}[1]{#1}
\csname url@samestyle\endcsname
\providecommand{\newblock}{\relax}
\providecommand{\bibinfo}[2]{#2}
\providecommand{\BIBentrySTDinterwordspacing}{\spaceskip=0pt\relax}
\providecommand{\BIBentryALTinterwordstretchfactor}{4}
\providecommand{\BIBentryALTinterwordspacing}{\spaceskip=\fontdimen2\font plus
\BIBentryALTinterwordstretchfactor\fontdimen3\font minus
  \fontdimen4\font\relax}
\providecommand{\BIBforeignlanguage}[2]{{%
\expandafter\ifx\csname l@#1\endcsname\relax
\typeout{** WARNING: IEEEtran.bst: No hyphenation pattern has been}%
\typeout{** loaded for the language `#1'. Using the pattern for}%
\typeout{** the default language instead.}%
\else
\language=\csname l@#1\endcsname
\fi
#2}}
\providecommand{\BIBdecl}{\relax}
\BIBdecl

\bibitem{javaid2018cm}
N.~Javaid, A.~Sher, H.~Nasir, and N.~Guizani, ``Intelligence in {IoT}-based
  {5G} networks: Opportunities and challenges,'' \emph{IEEE Communications
  Magazine}, vol.~56, no.~10, pp. 94--100, 2018.

\bibitem{sharm2020comst}
S.~K. Sharma and X.~Wang, ``Toward massive machine type communications in
  ultra-dense cellular {IoT} networks: Current issues and machine
  learning-assisted solutions,'' \emph{IEEE Communications Surveys \&
  Tutorials}, vol.~22, no.~1, pp. 426--471, 2020.

\bibitem{jian2021ojcoms}
W.~Jiang, B.~Han, M.~A. Habibi, and H.~D. Schotten, ``The road towards {6G: A}
  comprehensive survey,'' \emph{IEEE Open Journal of the Communications
  Society}, vol.~2, pp. 334--366, 2021.

\bibitem{vaezi2022comst}
M.~Vaezi, A.~Azari, S.~R. Khosravirad, M.~Shirvanimoghaddam, M.~M. Azari,
  D.~Chasaki, and P.~Popovski, ``Cellular, wide-area, and non-terrestrial
  {IoT}: A survey on {5G} advances and the road toward {6G},'' \emph{IEEE
  Communications Surveys \& Tutorials}, vol.~24, no.~2, pp. 1117--1174, 2022.

\bibitem{xia2020wcm}
T.~Xia, M.~M. Wang, J.~Zhang, and L.~Wang, ``Maritime internet of things:
  Challenges and solutions,'' \emph{IEEE Wireless Communications}, vol.~27,
  no.~2, pp. 188--196, 2020.

\bibitem{wang2020machine}
M.~M. Wang, J.~Zhang, and X.~You, ``Machine-type communication for maritime
  internet of things: {A} design,'' \emph{IEEE Communications Surveys \&
  Tutorials}, vol.~22, no.~4, pp. 2550--2585, 2020.

\bibitem{zeng2016cm}
Y.~Zeng, R.~Zhang, and T.~J. Lim, ``Wireless communications with unmanned
  aerial vehicles: opportunities and challenges,'' \emph{IEEE Communications
  Magazine}, vol.~54, no.~5, pp. 36--42, 2016.

\bibitem{Wang2021jsac}
Y.~Wang, W.~Feng, J.~Wang, and T.~Q.~S. Quek, ``Hybrid
  satellite-{UAV}-terrestrial networks for {6G} ubiquitous coverage: A maritime
  communications perspective,'' \emph{IEEE Journal on Selected Areas in
  Communications}, vol.~39, no.~11, pp. 3475--3490, 2021.

\bibitem{bithas2019sensors}
\BIBentryALTinterwordspacing
P.~S. Bithas, E.~T. Michailidis, N.~Nomikos, D.~Vouyioukas, and A.~G. Kanatas,
  ``A survey on machine-learning techniques for {UAV}-based communications,''
  \emph{Sensors}, vol.~19, no.~23, 2019. [Online]. Available:
  \url{https://www.mdpi.com/1424-8220/19/23/5170}
\BIBentrySTDinterwordspacing

\bibitem{jahanbakht2021comst}
M.~Jahanbakht, W.~Xiang, L.~Hanzo, and M.~Rahimi~Azghadi, ``Internet of
  underwater things and big marine data analytics-a comprehensive survey,''
  \emph{IEEE Communications Surveys \& Tutorials}, vol.~23, no.~2, pp.
  904--956, 2021.

\bibitem{guan2021cm}
S.~Guan, J.~Wang, C.~Jiang, R.~Duan, Y.~Ren, and T.~Q.~S. Quek, ``{MagicNet}:
  {T}he maritime giant cellular network,'' \emph{IEEE Communications Magazine},
  vol.~59, no.~3, pp. 117--123, 2021.

\bibitem{haidine2021agers}
A.~Haidine, A.~Aqqal, and A.~Dahbi, ``Communications backbone for environment
  monitoring applications in smart maritime ports-case study of a moroccan
  port,'' in \emph{2021 IEEE Asia-Pacific Conference on Geoscience, Electronics
  and Remote Sensing Technology (AGERS)}, 2021, pp. 136--140.

\bibitem{alqurashi2022arxiv}
\BIBentryALTinterwordspacing
F.~S. Alqurashi, A.~Trichili, N.~Saeed, B.~S. Ooi, and M.-S. Alouini,
  ``Maritime communications: A survey on enabling technologies, opportunities,
  and challenges,'' 2022. [Online]. Available:
  \url{https://arxiv.org/abs/2204.12824}
\BIBentrySTDinterwordspacing

\bibitem{matracia2022ojcoms}
M.~Matracia, N.~Saeed, M.~A. Kishk, and M.-S. Alouini, ``Post-disaster
  communications: Enabling technologies, architectures, and open challenges,''
  \emph{IEEE Open Journal of the Communications Society}, vol.~3, pp.
  1177--1205, 2022.

\bibitem{wei2021iotj}
T.~Wei, W.~Feng, Y.~Chen, C.-X. Wang, N.~Ge, and J.~Lu, ``Hybrid
  satellite-terrestrial communication networks for the maritime internet of
  things: Key technologies, opportunities, and challenges,'' \emph{IEEE
  Internet of Things Journal}, vol.~8, no.~11, pp. 8910--8934, 2021.

\bibitem{starlink}
``{Starlink Maritime},'' \url{https://https://www.starlink.com/maritime}, 2008,
  [Online; accessed September, 2022].

\bibitem{hyoungwon2017ict}
H.~Seo, J.~Shim, S.~Ha, Y.-S. Kim, and J.~Jeong, ``Ultra long range {LTE} ocean
  coverage solution,'' in \emph{2017 24th International Conference on
  Telecommunications (ICT)}, 2017, pp. 1--5.

\bibitem{huo2020iotj}
Y.~Huo, X.~Dong, and S.~Beatty, ``Cellular communications in ocean waves for
  maritime internet of things,'' \emph{IEEE Internet of Things Journal},
  vol.~7, no.~10, pp. 9965--9979, 2020.

\bibitem{bithas2020uav}
P.~S. Bithas, V.~Nikolaidis, A.~G. Kanatas, and G.~K. Karagiannidis,
  ``{UAV}-to-ground communications: {C}hannel modeling and {UAV} selection,''
  \emph{IEEE Transactions on Communications}, vol.~68, no.~8, pp. 5135--5144,
  2020.

\bibitem{9040264}
M.~Giordani, M.~Polese, M.~Mezzavilla, S.~Rangan, and M.~Zorzi, ``Toward {6G}
  networks: {U}se cases and technologies,'' \emph{IEEE Communications
  Magazine}, vol.~58, no.~3, pp. 55--61, 2020.

\bibitem{zhu2021iotj}
X.~Zhu and C.~Jiang, ``Integrated satellite-terrestrial networks toward {6G}:
  Architectures, applications, and challenges,'' \emph{IEEE Internet of Things
  Journal}, vol.~9, no.~1, pp. 437--461, 2022.

\bibitem{Li2020wc}
X.~Li, W.~Feng, J.~Wang, Y.~Chen, N.~Ge, and C.-X. Wang, ``Enabling {5G} on the
  ocean: A hybrid satellite-{UAV}-terrestrial network solution,'' \emph{IEEE
  Wireless Communications}, vol.~27, no.~6, pp. 116--121, 2020.

\bibitem{zhao2019vtm}
N.~{Zhao}, F.~R. {Yu}, L.~{Fan}, Y.~{Chen}, J.~{Tang}, A.~{Nallanathan}, and
  V.~C.~M. {Leung}, ``Caching unmanned aerial vehicle-enabled small-cell
  networks: {E}mploying energy-efficient methods that store and retrieve
  popular content,'' \emph{IEEE Vehicular Technology Magazine}, vol.~14, no.~1,
  pp. 71--79, 2019.

\bibitem{zolich2019irsj}
A.~Zolich, D.~Palma, K.~Kansanen, K.~Fj\o{}rtoft, J.~Sousa, K.~H. Johansson,
  Y.~Jiang, H.~Dong, and T.~A. Johansen, ``Survey on communication and networks
  for autonomous marine systems,'' \emph{Journal of Intelligent \& Robotic
  Systems}, vol.~95, pp. 789--813, 2022.

\bibitem{luo2022sensorsj}
H.~Luo, J.~Wang, F.~Bu, R.~Ruby, K.~Wu, and Z.~Guo, ``Recent progress of
  air/water cross-boundary communications for underwater sensor networks: A
  review,'' \emph{IEEE Sensors Journal}, vol.~22, no.~9, pp. 8360--8382, 2022.

\bibitem{wang2018access}
J.~Wang, H.~Zhou, Y.~Li, Q.~Sun, Y.~Wu, S.~Jin, T.~Q.~S. Quek, and C.~Xu,
  ``Wireless channel models for maritime communications,'' \emph{IEEE Access},
  vol.~6, pp. 68\,070--68\,088, 2018.

\bibitem{ai2020wcm}
B.~Ai, R.~He, H.~Zhang, M.~Yang, Z.~Ma, G.~Sun, and Z.~Zhong, ``Feeder
  communication for integrated networks,'' \emph{IEEE Wireless Communications},
  vol.~27, no.~6, pp. 20--27, 2020.

\bibitem{li2021tii}
Y.~Li, J.~Huang, Q.~Sun, T.~Sun, and S.~Wang, ``Cognitive service architecture
  for {6G} core network,'' \emph{IEEE Transactions on Industrial Informatics},
  vol.~17, no.~10, pp. 7193--7203, 2021.

\bibitem{aitallal2022ai2sd}
A.~Ait~Allal, L.~E. Amrani, A.~Haidine, K.~Mansouri, and M.~Youssfi, ``Towards
  deployment of {UAV}'s for autonomous ships' {5G} mobile communication,'' in
  \emph{Advanced Intelligent Systems for Sustainable Development (AI2SD'2020)},
  J.~Kacprzyk, V.~E. Balas, and M.~Ezziyyani, Eds.\hskip 1em plus 0.5em minus
  0.4em\relax Cham: Springer International Publishing, 2022, pp. 944--956.

\bibitem{zhong2019imm}
M.~Zhong, Y.~Yang, H.~Yao, X.~Fu, O.~A. Dobre, and O.~Postolache, ``{5G} and
  {IoT}: Towards a new era of communications and measurements,'' \emph{IEEE
  Instrumentation \& Measurement Magazine}, vol.~22, no.~6, pp. 18--26, 2019.

\bibitem{teixera2020access}
F.~B. Teixeira, R.~Campos, and M.~Ricardo, ``Height optimization in aerial
  networks for enhanced broadband communications at sea,'' \emph{IEEE Access},
  vol.~8, pp. 28\,311--28\,323, 2020.

\bibitem{park2021milcom}
J.~Y. Park, F.~S. Encarnado, C.~R. Flint, and K.~H. Redwine, ``Characterization
  and modeling of frequency-selective ship-to-ship channels,'' in \emph{MILCOM
  2021 - 2021 IEEE Military Communications Conference (MILCOM)}, 2021, pp.
  550--555.

\bibitem{liu2019tcc}
C.~Liu, Y.~Li, R.~Jiang, F.~Hong, and Z.~Guo, ``Oceannet: a low-cost
  large-scale maritime communication architecture based on {D2D} communication
  technology,'' \emph{Proceedings of the ACM Turing Celebration Conference -
  China}, 2019.

\bibitem{ramezani2022network}
P.~Ramezani, B.~Lyu, and A.~Jamalipour, ``Toward {RIS}-enhanced integrated
  terrestrial/non-terrestrial connectivity in {6G},'' \emph{IEEE Network}, pp.
  1--9, 2022.

\bibitem{kaushal2017comst}
H.~Kaushal and G.~Kaddoum, ``Optical communication in space: {C}hallenges and
  mitigation techniques,'' \emph{IEEE Communications Surveys \& Tutorials},
  vol.~19, no.~1, pp. 57--96, 2017.

\bibitem{lionis2021optics}
\BIBentryALTinterwordspacing
A.~Lionis, K.~Peppas, H.~E. Nistazakis, A.~Tsigopoulos, and K.~Cohn,
  ``Statistical modeling of received signal strength for an {FSO} link over
  maritime environment,'' \emph{Optics Communications}, vol. 489, p. 126858,
  2021. [Online]. Available:
  \url{https://www.sciencedirect.com/science/article/pii/S0030401821001085}
\BIBentrySTDinterwordspacing

\bibitem{jamali2016twc}
V.~Jamali, D.~S. Michalopoulos, M.~Uysal, and R.~Schober, ``Link allocation for
  multiuser systems with hybrid {RF/FSO} backhaul: Delay-limited and
  delay-tolerant designs,'' \emph{IEEE Transactions on Wireless
  Communications}, vol.~15, no.~5, pp. 3281--3295, 2016.

\bibitem{arienzo2019tccn}
L.~Arienzo, ``Green {RF/FSO} communications in cognitive relay-based space
  information networks for maritime surveillance,'' \emph{IEEE Transactions on
  Cognitive Communications and Networking}, vol.~5, no.~4, pp. 1182--1193,
  2019.

\bibitem{upadhya2020tcom}
A.~Upadhya, V.~K. Dwivedi, and G.~K. Karagiannidis, ``On the effect of
  interference and misalignment error in mixed {RF/FSO} systems over
  generalized fading channels,'' \emph{IEEE Transactions on Communications},
  vol.~68, no.~6, pp. 3681--3695, 2020.

\bibitem{ninos2021ojcoms}
M.~P. Ninos, P.~Mukherjee, C.~Psomas, and I.~Krikidis, ``Full-duplex {DF}
  relaying with parallel hybrid {FSO/RF} transmissions,'' \emph{IEEE Open
  Journal of the Communications Society}, vol.~2, pp. 2502--2515, 2021.

\bibitem{huang2020jlt}
G.~Huang, L.~Zhang, Y.~Jiang, and Z.~Wu, ``A general orthogonal transform aided
  {MIMO} design for reliable maritime visible light communications,''
  \emph{Journal of Lightwave Technology}, vol.~38, no.~23, pp. 6549--6560,
  2020.

\bibitem{eso2021ptl}
E.~Eso, Z.~Ghassemlooy, S.~Zvanovec, P.~Pesek, and J.~Sathian,
  ``Vehicle-to-vehicle relay-assisted {VLC} with misalignment induced azimuth
  or elevation offset angles,'' \emph{IEEE Photonics Technology Letters},
  vol.~33, no.~16, pp. 908--911, 2021.

\bibitem{jiang2018comst1}
S.~Jiang, ``On reliable data transfer in underwater acoustic networks: A survey
  from networking perspective,'' \emph{IEEE Communications Surveys \&
  Tutorials}, vol.~20, no.~2, pp. 1036--1055, 2018.

\bibitem{jiang2018comst2}
------, ``State-of-the-art medium access control {(MAC)} protocols for
  underwater acoustic networks: {A} survey based on a {MAC} reference model,''
  \emph{IEEE Communications Surveys \& Tutorials}, vol.~20, no.~1, pp. 96--131,
  2018.

\bibitem{luo2021comst}
J.~Luo, Y.~Chen, M.~Wu, and Y.~Yang, ``A survey of routing protocols for
  underwater wireless sensor networks,'' \emph{IEEE Communications Surveys \&
  Tutorials}, vol.~23, no.~1, pp. 137--160, 2021.

\bibitem{luo2019access}
H.~Luo, X.~Xie, G.~Han, R.~Ruby, F.~Hong, and Y.~Liang, ``Multimodal
  acoustic-{RF} adaptive routing protocols for underwater wireless sensor
  networks,'' \emph{IEEE Access}, vol.~7, pp. 134\,954--134\,967, 2019.

\bibitem{yang2020network}
T.~Yang, J.~Chen, and N.~Zhang, ``{AI}-empowered maritime internet of things:
  {A} parallel-network-driven approach,'' \emph{IEEE Network}, vol.~34, no.~5,
  pp. 54--59, 2020.

\bibitem{nomikos2022ojcoms}
N.~Nomikos, M.~S. Talebi, T.~Charalambous, and R.~Wichman, ``Bandit-based power
  control in full-duplex cooperative relay networks with strict-sense
  stationary and non-stationary wireless communication channels,'' \emph{IEEE
  Open Journal of the Communications Society}, vol.~3, pp. 366--378, 2022.

\bibitem{mukherjee2020tsc}
A.~Mukherjee, S.~Misra, V.~S.~P. Chandra, and N.~S. Raghuwanshi, ``{ECoR:}
  energy-aware collaborative routing for task offload in sustainable {UAV}
  swarms,'' \emph{IEEE Transactions on Sustainable Computing}, vol.~5, no.~4,
  pp. 514--525, 2020.

\bibitem{hu2021tvt}
Y.~Hu, X.~Yuan, G.~Zhang, and A.~Schmeink, ``Sustainable wireless sensor
  networks with {UAV}-enabled wireless power transfer,'' \emph{IEEE
  Transactions on Vehicular Technology}, vol.~70, no.~8, pp. 8050--8064, 2021.

\bibitem{muhammad2021tvt}
S.~Muhammad~Hashir, A.~Mehrabi, M.~R. Mili, M.~J. Emadi, D.~W.~K. Ng, and
  I.~Krikidis, ``Performance trade-off in {UAV}-aided wireless-powered
  communication networks via multi-objective optimization,'' \emph{IEEE
  Transactions on Vehicular Technology}, vol.~70, no.~12, pp. 13\,430--13\,435,
  2021.

\bibitem{nomikos2020vehcom}
\BIBentryALTinterwordspacing
N.~Nomikos, E.~T. Michailidis, P.~Trakadas, D.~Vouyioukas, H.~Karl, J.~Martrat,
  T.~Zahariadis, K.~Papadopoulos, and S.~Voliotis, ``A {UAV}-based moving {5G
  RAN} for massive connectivity of mobile users and {IoT} devices,''
  \emph{Vehicular Communications}, vol.~25, p. 100250, 2020. [Online].
  Available:
  \url{https://www.sciencedirect.com/science/article/pii/S2214209620300218}
\BIBentrySTDinterwordspacing

\bibitem{feng2022ciot}
H.~Feng, Z.~Cui, and T.~Yang, ``Cache optimization strategy for mobile edge
  computing in maritime {IoT},'' in \emph{2022 5th Conference on Cloud and
  Internet of Things (CIoT)}, 2022, pp. 213--219.

\bibitem{Timmins2009VEH}
I.~J. Timmins and S.~O'Young, ``Marine communications channel modeling using
  the finite-difference time domain method,'' \emph{IEEE Transactions on
  Vehicular Technology}, vol.~58, no.~6, pp. 2626--2637, 2009.

\bibitem{Liu2021JSAC}
Y.~Liu, C.-X. Wang, H.~Chang, Y.~He, and J.~Bian, ``A novel non-stationary {6G
  UAV} channel model for maritime communications,'' \emph{IEEE Journal on
  Selected Areas in Communications}, vol.~39, no.~10, pp. 2992--3005, 2021.

\bibitem{Gao2020China}
Z.~Gao, B.~Liu, Z.~Cheng, C.~Chen, and L.~Huang, ``Marine mobile wireless
  channel modeling based on improved spatial partitioning ray tracing,''
  \emph{China Communications}, vol.~17, no.~3, pp. 1--11, 2020.

\bibitem{Rasheed2022ieee_trans}
I.~Rasheed, M.~Asif, A.~Ihsan, W.~U. Khan, M.~Ahmed, and K.~M. Rabie,
  ``{LSTM}-based distributed conditional generative adversarial network for
  data-driven {5G}-enabled maritime {UAV} communications,'' \emph{IEEE
  Transactions on Intelligent Transportation Systems}, pp. 1--16, 2022.

\bibitem{Liu2021JCN}
F.~Liu, J.~Pan, X.~Zhou, and G.~Y. Li, ``Atmospheric ducting effect in wireless
  communications: Challenges and opportunities,'' \emph{Journal of
  Communications and Information Networks}, vol.~6, no.~2, pp. 101--109, 2021.

\bibitem{Cao2021}
X.~Cao, S.~Yan, and M.~Peng, ``Mmwave-based beamforming for capacity
  maximization in {UAV}-aided maritime communication networks,'' in \emph{2021
  IEEE/CIC International Conference on Communications in China (ICCC)}, 2021,
  pp. 451--456.

\bibitem{Wang2020}
Y.~Wang, X.~Fang, W.~Feng, Y.~Chen, N.~Ge, and Z.~Lu, ``On-demand coverage for
  maritime hybrid satellite-{UAV}-terrestrial networks,'' in \emph{2020
  International Conference on Wireless Communications and Signal Processing
  (WCSP)}, 2020, pp. 483--488.

\bibitem{Ghanbari2022}
M.~Ghanbari, M.~Ataee, and S.~M. Sajad~Sadough, ``Outage performance analysis
  for {UAV}-based mixed underwater-{FSO} communication under pointing errors,''
  in \emph{2022 4th West Asian Symposium on Optical and Millimeter-wave
  Wireless Communications (WASOWC)}, 2022, pp. 1--5.

\bibitem{Wang2020ieee_internet}
Q.~Wang, H.-N. Dai, Q.~Wang, M.~K. Shukla, W.~Zhang, and C.~G. Soares, ``On
  connectivity of {UAV}-assisted data acquisition for underwater internet of
  things,'' \emph{IEEE Internet of Things Journal}, vol.~7, no.~6, pp.
  5371--5385, 2020.

\bibitem{Fang2020}
X.~Fang, Y.~Wang, W.~Feng, Y.~Chen, and B.~Ai, ``Power allocation for maritime
  cognitive satellite-{UAV}-terrestrial networks,'' in \emph{2020 IEEE 19th
  International Conference on Cognitive Informatics \& Cognitive Computing
  (ICCI*CC)}, 2020, pp. 139--143.

\bibitem{Liu2021conf}
K.~Liu, P.~Li, C.~Liu, L.~Xiao, and L.~Jia, ``{UAV}-aided anti-jamming maritime
  communications: {A} deep reinforcement learning approach,'' in \emph{2021
  13th International Conference on Wireless Communications and Signal
  Processing (WCSP)}, 2021, pp. 1--6.

\bibitem{Rahimi2022}
P.~Rahimi, C.~Chrysostomou, I.~Kyriakides, and V.~Vassiliou, ``An
  energy-efficient machine-type communication for maritime internet of
  things,'' in \emph{2020 11th IEEE Annual Information Technology, Electronics
  and Mobile Communication Conference (IEMCON)}, 2020, pp. 0668--0676.

\bibitem{Li2020ieeetrans}
X.~Li, W.~Feng, Y.~Chen, C.-X. Wang, and N.~Ge, ``Maritime coverage enhancement
  using {UAVs} coordinated with hybrid satellite-terrestrial networks,''
  \emph{IEEE Transactions on Communications}, vol.~68, no.~4, pp. 2355--2369,
  2020.

\bibitem{Hong2021ieeenetwork}
T.~Hong, M.~Lv, S.~Zheng, and H.~Hong, ``Key technologies in {6G SAGS IoT}:
  Shape-adaptive antenna and radar-communication integration,'' \emph{IEEE
  Network}, vol.~35, no.~5, pp. 150--157, 2021.

\bibitem{Pottoo2022}
S.~Nazir~Pottoo, P.~Gunnar~Ellingsen, and T.~Dac~Ho, ``Evaluations of
  {UAV}-enabled {FSO} communications in the arctic,'' in \emph{2022 IEEE/SICE
  International Symposium on System Integration (SII)}, 2022, pp. 297--302.

\bibitem{Ma2020network}
R.~Ma, R.~Wang, G.~Liu, H.-H. Chen, and Z.~Qin, ``{UAV}-assisted data
  collection for ocean monitoring networks,'' \emph{IEEE Network}, vol.~34,
  no.~6, pp. 250--258, 2020.

\bibitem{Chen2020MDPI}
\BIBentryALTinterwordspacing
H.~Chen, F.~Yin, W.~Huang, M.~Liu, and D.~Li, ``Ocean surface drifting buoy
  system based on {UAV}-enabled wireless powered relay network,''
  \emph{Sensors}, vol.~20, no.~9, 2020. [Online]. Available:
  \url{https://www.mdpi.com/1424-8220/20/9/2598}
\BIBentrySTDinterwordspacing

\bibitem{Che2021conf}
Y.~Che, B.~Lin, and L.~Liu, ``Offshore container data transmission based on
  {UAV}-assisted relay,'' in \emph{2021 13th International Conference on
  Wireless Communications and Signal Processing (WCSP)}, 2021, pp. 1--5.

\bibitem{Kavuri2020IEEE}
S.~Kavuri, D.~Moltchanov, A.~Ometov, S.~Andreev, and Y.~Koucheryavy,
  ``Performance analysis of onshore {NB-IoT} for container tracking during
  near-the-shore vessel navigation,'' \emph{IEEE Internet of Things Journal},
  vol.~7, no.~4, pp. 2928--2943, 2020.

\bibitem{LyuSpringer}
L.~Lyu, Z.~Chu, and B.~Lin, ``Joint association and power optimization for
  multi-{UAV} assisted cooperative transmission in marine {IoT} networks,''
  \emph{IEEE Internet of Things Journal}, vol.~14, pp. 3307--3318, 2021.

\bibitem{Liu2021conf2}
L.~Liu, B.~Lin, and Y.~Che, ``Joint {UAV}-{BS} deployment and power allocation
  for maritime emergency communication system,'' in \emph{2021 13th
  International Conference on Wireless Communications and Signal Processing
  (WCSP)}, 2021, pp. 1--5.

\bibitem{Jia2021jsac}
Z.~Jia, M.~Sheng, J.~Li, D.~Zhou, and Z.~Han, ``Joint {HAP} access and {LEO}
  satellite backhaul in {6G}: Matching game-based approaches,'' \emph{IEEE
  Journal on Selected Areas in Communications}, vol.~39, no.~4, pp. 1147--1159,
  2021.

\bibitem{Cao2020conf}
H.~Cao, T.~Yang, Z.~Yin, X.~Sun, and D.~Li, ``Topological optimization
  algorithm for {HAP} assisted multi-unmanned ships communication,'' in
  \emph{2020 IEEE 92nd Vehicular Technology Conference (VTC2020-Fall)}, 2020,
  pp. 1--5.

\bibitem{Fang2022trans}
X.~Fang, W.~Feng, Y.~Wang, Y.~Chen, N.~Ge, Z.~Ding, and H.~Zhu, ``{NOMA}-based
  hybrid satellite-{UAV}-terrestrial networks for {6G} maritime coverage,''
  \emph{IEEE Transactions on Wireless Communications}, pp. 1--1, 2022.

\bibitem{Tang2021China}
R.~Tang, W.~Feng, Y.~Chen, and N.~Ge, ``{NOMA}-based {UAV} communications for
  maritime coverage enhancement,'' \emph{China Communications}, vol.~18, no.~4,
  pp. 230--243, 2021.

\bibitem{Ma2021internet}
R.~Ma, R.~Wang, G.~Liu, W.~Meng, and X.~Liu, ``{UAV}-aided cooperative data
  collection scheme for ocean monitoring networks,'' \emph{IEEE Internet of
  Things Journal}, vol.~8, no.~17, pp. 13\,222--13\,236, 2021.

\bibitem{Xie2018conf}
P.~Xie, ``An enhanced {OLSR} routing protocol based on node link expiration
  time and residual energy in ocean {FANETS},'' in \emph{2018 24th Asia-Pacific
  Conference on Communications (APCC)}, 2018, pp. 598--603.

\bibitem{Chaudhry2021trans}
S.~A. Chaudhry, A.~Irshad, M.~A. Khan, S.~A. Khan, S.~Nosheen, A.~A. AlZubi,
  and Y.~B. Zikria, ``A lightweight authentication scheme for {6G-IoT} enabled
  maritime transport system,'' \emph{IEEE Transactions on Intelligent
  Transportation Systems}, pp. 1--10, 2021.

\bibitem{security_elsevier}
M.~A. Khan, B.~A. Alzahrani, A.~Barnawi, A.~Al-Barakati, A.~Irshad, and S.~A.
  Chaudhry, ``A resource friendly authentication scheme for
  space-air-ground-sea integrated maritime communication network,'' \emph{Ocean
  Engineering}, vol. 250, p. 110894, 2022.

\bibitem{liu2021uav}
J.~Liu, Z.~Su, and Q.~Xu, ``{UAV-USV} cooperative task allocation for smart
  ocean networks,'' in \emph{2021 IEEE 23rd Int Conf on High Performance
  Computing \& Communications; 7th Int Conf on Data Science \& Systems; 19th
  Int Conf on Smart City; 7th Int Conf on Dependability in Sensor, Cloud \& Big
  Data Systems \& Application (HPCC/DSS/SmartCity/DependSys)}.\hskip 1em plus
  0.5em minus 0.4em\relax IEEE, 2021, pp. 1815--1820.

\bibitem{Nomikos2019ieee_access}
N.~Nomikos, E.~T. Michailidis, P.~Trakadas, D.~Vouyioukas, T.~Zahariadis, and
  I.~Krikidis, ``Flex-{NOMA}: Exploiting buffer-aided relay selection for
  massive connectivity in the {5G} uplink,'' \emph{IEEE Access}, vol.~7, pp.
  88\,743--88\,755, 2019.

\bibitem{feng2013jsac}
W.~Feng, Y.~Wang, N.~Ge, J.~Lu, and J.~Zhang, ``Virtual {MIMO} in multi-cell
  distributed antenna systems: {C}oordinated transmissions with large-scale
  {CSIT},'' \emph{IEEE Journal on Selected Areas in Communications}, vol.~31,
  no.~10, pp. 2067--2081, 2013.

\bibitem{choi2007twc}
W.~Choi and J.~G. Andrews, ``Downlink performance and capacity of distributed
  antenna systems in a multicell environment,'' \emph{IEEE Transactions on
  Wireless Communications}, vol.~6, no.~1, pp. 69--73, 2007.

\bibitem{Carrillo2021ieee_wirel_comm}
D.~Carrillo, K.~Mikhaylov, P.~J. Nardelli, S.~Andreev, and D.~B. da~Costa,
  ``Understanding {UAV}-based {WPCN}-aided capabilities for offshore monitoring
  applications,'' \emph{IEEE Wireless Communications}, vol.~28, no.~2, pp.
  114--120, 2021.

\bibitem{dai2022uav}
Y.~Dai, B.~Lin, Y.~Che, and L.~Lyu, ``{UAV}-assisted data offloading for smart
  container in offshore maritime communications,'' \emph{China Communications},
  vol.~19, no.~1, pp. 153--165, 2022.

\bibitem{hassan2021demand}
S.~S. Hassan, Y.~M. Park, and C.~S. Hong, ``On-demand {MEC} empowered {UAV}
  deployment for {6G} time-sensitive maritime internet of things,'' in
  \emph{2021 22nd Asia-Pacific Network Operations and Management Symposium
  (APNOMS)}.\hskip 1em plus 0.5em minus 0.4em\relax IEEE, 2021, pp. 386--389.

\bibitem{yang2022multi}
T.~Yang, S.~Gao, J.~Li, M.~Qin, X.~Sun, R.~Zhang, M.~Wang, and X.~Li,
  ``Multi-armed bandits learning for task offloading in maritime edge
  intelligence networks,'' \emph{IEEE Transactions on Vehicular Technology},
  vol.~71, no.~4, pp. 4212--4224, 2022.

\bibitem{zeng2020mobile}
J.~Zeng, J.~Sun, B.~Wu, and X.~Su, ``Mobile edge communications, computing, and
  caching {(MEC3)} technology in the maritime communication network,''
  \emph{China Communications}, vol.~17, no.~5, pp. 223--234, 2020.

\bibitem{xu2020deep}
F.~Xu, F.~Yang, C.~Zhao, and S.~Wu, ``Deep reinforcement learning based joint
  edge resource management in maritime network,'' \emph{China Communications},
  vol.~17, no.~5, pp. 211--222, 2020.

\bibitem{9678008}
Y.~Liu, J.~Yan, and X.~Zhao, ``Deep reinforcement learning based latency
  minimization for mobile edge computing with virtualization in maritime {UAV}
  communication network,'' \emph{IEEE Transactions on Vehicular Technology},
  vol.~71, no.~4, pp. 4225--4236, 2022.

\bibitem{zeng2022collaborative}
H.~Zeng, R.~Li, Z.~Su, Q.~Xu, Y.~Wang, M.~Dai, T.~H. Luan, X.~Sun, and D.~Liu,
  ``Collaborative computation offloading for {UAV}s and {USV} fleets in
  communication networks,'' in \emph{2022 International Wireless Communications
  and Mobile Computing (IWCMC)}.\hskip 1em plus 0.5em minus 0.4em\relax IEEE,
  2022, pp. 949--954.

\bibitem{hassan2021blue}
S.~S. Hassan, Y.~K. Tun, W.~Saad, Z.~Han, and C.~S. Hong, ``Blue data
  computation maximization in {6G} space-air-sea non-terrestrial networks,'' in
  \emph{2021 IEEE Global Communications Conference (GLOBECOM)}.\hskip 1em plus
  0.5em minus 0.4em\relax IEEE, 2021, pp. 1--6.

\bibitem{dai2022deep}
Y.~Dai, Z.~Liang, L.~Lyu, and B.~Lin, ``Deep reinforcement learning-based {UAV}
  data collection and offloading in {NOMA}-enabled marine {IoT} systems,''
  \emph{Wireless Communications and Mobile Computing}, vol. 2022, 2022.

\bibitem{gao2020multi}
S.~Gao, T.~Yang, H.~Ni, and G.~Zhang, ``Multi-armed bandits scheme for tasks
  offloading in {MEC}-enabled maritime communication networks,'' in \emph{2020
  IEEE/CIC International Conference on Communications in China (ICCC)}.\hskip
  1em plus 0.5em minus 0.4em\relax IEEE, 2020, pp. 232--237.

\bibitem{pang2020space}
Y.~Pang, D.~Wang, D.~Wang, L.~Guan, C.~Zhang, and M.~Zhang, ``A
  space-air-ground integrated network assisted maritime communication network
  based on mobile edge computing,'' in \emph{2020 IEEE World Congress on
  Services (SERVICES)}.\hskip 1em plus 0.5em minus 0.4em\relax IEEE, 2020, pp.
  269--274.

\bibitem{Velascodronesmdpi}
\BIBentryALTinterwordspacing
O.~Velasco, J.~Valente, P.~J. Alhama~Blanco, and M.~Abderrahim, ``An open
  simulation strategy for rapid control design in aerial and maritime drone
  teams: {A} comprehensive tutorial,'' \emph{Drones}, vol.~4, no.~3, 2020.
  [Online]. Available: \url{https://www.mdpi.com/2504-446X/4/3/37}
\BIBentrySTDinterwordspacing

\bibitem{Zhang2021conf}
Y.~Zhang, B.~Lin, X.~Hu, and Z.~Wang, ``Deployment and optimization of
  multi-{UAV}-assisted maritime internet of things for waterway data
  collection,'' in \emph{2021 International Conference on Security, Pattern
  Analysis, and Cybernetics (SPAC)}, 2021, pp. 577--580.

\bibitem{Guan2021conf}
S.~Guan, J.~Wang, C.~Jiang, X.~Hou, Z.~Fang, and Y.~Ren, ``Efficient on-demand
  {UAV} deployment and configuration for off-shore relay communications,'' in
  \emph{2021 International Wireless Communications and Mobile Computing
  (IWCMC)}, 2021, pp. 997--1002.

\bibitem{Zhang2020confglobecom}
Y.~Zhang, J.~Lyu, and L.~Fu, ``Energy-efficient cyclical trajectory design for
  {UAV}-aided maritime data collection in wind,'' in \emph{GLOBECOM 2020 - 2020
  IEEE Global Communications Conference}, 2020, pp. 1--6.

\bibitem{Zhang2022trans}
------, ``Energy-efficient trajectory design for {UAV}-aided maritime data
  collection in wind,'' \emph{IEEE Transactions on Wireless Communications},
  pp. 1--1, 2022.

\bibitem{Liu2022}
Z.~Liu, X.~Meng, Y.~Yang, K.~Ma, and X.~Guan, ``Energy-efficient {UAV}-aided
  ocean monitoring networks: Joint resource allocation and trajectory design,''
  \emph{IEEE Internet of Things Journal}, vol.~9, no.~18, pp. 17\,871--17\,884,
  2022.

\bibitem{Lyu2022conf}
L.~Lyu, Z.~Chu, B.~Lin, Y.~Dai, and N.~Cheng, ``Fast trajectory planning for
  {UAV}-enabled maritime {IoT} systems: {A} fermat-point based approach,''
  \emph{IEEE Wireless Communications Letters}, vol.~11, no.~2, pp. 328--332,
  2022.

\bibitem{Li2021conf}
H.~Li, C.~Yu, C.~Zhang, H.~Jiao, B.~Lin, and R.~He, ``Maritime multi-relay
  communications based on {UAV} trajectory adjustment and dual {Q}-learning,''
  in \emph{2021 International Conference on Security, Pattern Analysis, and
  Cybernetics (SPAC)}, 2021, pp. 571--576.

\bibitem{Li2019conference}
X.~Li, W.~Feng, Y.~Chen, C.-X. Wang, and N.~Ge, ``{UAV}-enabled accompanying
  coverage for hybrid satellite-{UAV}-terrestrial maritime communications,'' in
  \emph{2019 28th Wireless and Optical Communications Conference (WOCC)}, 2019,
  pp. 1--5.

\bibitem{Zhang2020chinacommunications}
J.~Zhang, F.~Liang, B.~Li, Z.~Yang, Y.~Wu, and H.~Zhu, ``Placement optimization
  of caching {UAV}-assisted mobile relay maritime communication,'' \emph{China
  Communications}, vol.~17, no.~8, pp. 209--219, 2020.

\bibitem{Ho2021conf}
T.~Dac~Ho, E.~Ingar~Grøtli, and T.~Arne~Johansen, ``{PSO} and kalman
  filter-based node motion prediction for data collection from ocean wireless
  sensors network with {UAV},'' in \emph{2021 IEEE International Conference on
  Consumer Electronics (ICCE)}, 2021, pp. 1--7.

\bibitem{Yan2021}
M.~Yan, H.~Yuan, and J.~e.~a. Xu, ``Task allocation and route planning of
  multiple {UAV}s in a marine environment based on an improved particle swarm
  optimization algorithm,'' in \emph{EURASIP J. Adv. Signal Process}, vol.~94,
  2021.

\bibitem{Kim2018}
H.~Kim, L.~Mokdad, and J.~Ben-Othman, ``Designing {UAV} surveillance frameworks
  for smart city and extensive ocean with differential perspectives,''
  \emph{IEEE Communications Magazine}, vol.~56, no.~4, pp. 98--104, 2018.

\bibitem{ZuoBeidou2020}
J.~Zuo, J.~Chen, Z.~Li, Z.~Li, Z.~Liu, and Z.~Han, ``Research on maritime
  rescue {UAV} based on beidou {CNSS} and extended square search algorithm,''
  in \emph{2020 International Conference on Communications, Information System
  and Computer Engineering (CISCE)}, 2020, pp. 102--106.

\bibitem{Oliveira2019conf}
T.~Oliveira, A.~Agamyrzyansc, and L.~M. Correia, ``A computational tool to
  assess communications' range and capacity limits of ad-hoc networks of {UAV}s
  operating in maritime scenarios,'' in \emph{2019 International Conference on
  Unmanned Aircraft Systems (ICUAS)}, 2019, pp. 246--252.

\bibitem{Yang2020}
T.~Yang, Z.~Jiang, R.~Sun, N.~Cheng, and H.~Feng, ``Maritime search and rescue
  based on group mobile computing for unmanned aerial vehicles and unmanned
  surface vehicles,'' \emph{IEEE Transactions on Industrial Informatics},
  vol.~16, no.~12, pp. 7700--7708, 2020.

\bibitem{Wu2020}
Y.~Wu, K.~H. Low, and C.~Lv, ``Cooperative path planning for heterogeneous
  unmanned vehicles in a search-and-track mission aiming at an underwater
  target,'' \emph{IEEE Transactions on Vehicular Technology}, vol.~69, no.~6,
  pp. 6782--6787, 2020.

\bibitem{Brown2020}
A.~Brown and D.~Anderson, ``Trajectory optimization for high-altitude
  long-endurance {UAV} maritime radar surveillance,'' \emph{IEEE Transactions
  on Aerospace and Electronic Systems}, vol.~56, no.~3, pp. 2406--2421, 2020.

\bibitem{Sollesnes2018conf}
E.~Sollesnes, O.~M. Brokstad, R.~K. boe, B.~Vågen, A.~Carella, A.~Alcocer,
  A.~P. Zolich, and T.~A. Johansen, ``Towards autonomous ocean observing
  systems using miniature underwater gliders with {UAV} deployment and recovery
  capabilities,'' in \emph{2018 IEEE/OES Autonomous Underwater Vehicle Workshop
  (AUV)}, 2018, pp. 1--5.

\bibitem{Yokota2021mdpi}
\BIBentryALTinterwordspacing
Y.~Yokota and T.~Matsuda, ``Underwater communication using {UAV}s to realize
  high-speed {AUV} deployment,'' \emph{Remote Sensing}, vol.~13, no.~20, 2021.
  [Online]. Available: \url{https://www.mdpi.com/2072-4292/13/20/4173}
\BIBentrySTDinterwordspacing

\bibitem{Yu2022conf}
Y.~Yu, J.~Rodr\'{i}guez-Pi\~{n}eiro, X.~Shunqin, Y.~Tong, J.~Zhang, and X.~Yin,
  ``Measurement-based propagation channel modeling for communication between a
  {UAV and a USV},'' in \emph{2022 16th European Conference on Antennas and
  Propagation (EuCAP)}, 2022, pp. 01--05.

\bibitem{Pokorny2021sensors}
\BIBentryALTinterwordspacing
J.~Pokorny, K.~Ma, S.~Saafi, J.~Frolka, J.~Villa, M.~Gerasimenko,
  Y.~Koucheryavy, and J.~Hosek, ``Prototype design and experimental evaluation
  of autonomous collaborative communication system for emerging maritime use
  cases,'' \emph{Sensors}, vol.~21, no.~11, 2021. [Online]. Available:
  \url{https://www.mdpi.com/1424-8220/21/11/3871}
\BIBentrySTDinterwordspacing

\bibitem{Teixeira2020ieeeaccess}
F.~B. Teixeira, R.~Campos, and M.~Ricardo, ``Height optimization in aerial
  networks for enhanced broadband communications at sea,'' \emph{IEEE Access},
  vol.~8, pp. 28\,311--28\,323, 2020.

\bibitem{Guldenring2019}
J.~G\"{u}ldenring, L.~Koring, P.~Gorczak, and C.~Wietfeld, ``Heterogeneous
  multilink aggregation for reliable {UAV} communication in maritime search and
  rescue missions,'' in \emph{2019 International Conference on Wireless and
  Mobile Computing, Networking and Communications (WiMob)}, 2019, pp. 215--220.

\bibitem{Gorczak2019}
P.~Gorczak, C.~Bektas, F.~Kurtz, T.~Lübcke, and C.~Wietfeld, ``Robust cellular
  communications for unmanned aerial vehicles in maritime search and rescue,''
  in \emph{2019 IEEE International Symposium on Safety, Security, and Rescue
  Robotics (SSRR)}, 2019, pp. 229--234.

\bibitem{drones4020016}
\BIBentryALTinterwordspacing
J.~Güldenring, P.~Gorczak, F.~Eckermann, M.~Patchou, J.~Tiemann, F.~Kurtz, and
  C.~Wietfeld, ``Reliable long-range multi-link communication for unmanned
  search and rescue aircraft systems in beyond visual line of sight
  operation,'' \emph{Drones}, vol.~4, no.~2, 2020. [Online]. Available:
  \url{https://www.mdpi.com/2504-446X/4/2/16}
\BIBentrySTDinterwordspacing

\bibitem{Ribeiro2018}
M.~Ribeiro, J.~Galante, J.~Teixeira, and J.~B. de~Sousa, ``Use of a {UAV} as an
  acoustic communication relay system,'' in \emph{2018 International Conference
  on Unmanned Aircraft Systems (ICUAS)}, 2018, pp. 788--795.

\bibitem{Zolich2017}
A.~Zolich, A.~Sœgrov, E.~Vågsholm, V.~Hovstein, and T.~A. Johansen,
  ``Coordinated maritime missions of unmanned vehicles — network architecture
  and performance analysis,'' in \emph{2017 IEEE International Conference on
  Communications (ICC)}, 2017, pp. 1--7.

\bibitem{Elhoushy2022}
S.~Elhoushy, M.~Ibrahim, and W.~Hamouda, ``Cell-free massive {MIMO}: A
  survey,'' \emph{IEEE Communications Surveys \& Tutorials}, vol.~24, no.~1,
  pp. 492--523, 2022.

\bibitem{iccdmimo}
D.~Kudathanthirige and G.~Amarasuriya, ``Distributed massive {MIMO} downlink,''
  in \emph{ICC 2019 - 2019 IEEE International Conference on Communications
  (ICC)}, 2019, pp. 1--7.

\bibitem{nguyen2022tccn}
M.-H.~T. Nguyen, E.~Garcia-Palacios, T.~Do-Duy, O.~A. Dobre, and T.~Q. Duong,
  ``{UAV}-aided aerial reconfigurable intelligent surface communications with
  massive {MIMO} system,'' \emph{IEEE Transactions on Cognitive Communications
  and Networking}, pp. 1--1, 2022.

\bibitem{shang2021wcm}
B.~Shang, R.~Shafin, and L.~Liu, ``{UAV} swarm-enabled aerial reconfigurable
  intelligent surface {(SARIS)},'' \emph{IEEE Wireless Communications},
  vol.~28, no.~5, pp. 156--163, 2021.

\bibitem{liu2022jsac}
Y.~Liu, S.~Zhang, X.~Mu, Z.~Ding, R.~Schober, N.~Al-Dhahir, E.~Hossain, and
  X.~Shen, ``Evolution of {NOMA} toward next generation multiple access
  {(NGMA)} for {6G},'' \emph{IEEE Journal on Selected Areas in Communications},
  vol.~40, no.~4, pp. 1037--1071, 2022.

\bibitem{ding2018tcom}
Z.~{Ding}, P.~{Fan}, G.~K. {Karagiannidis}, R.~{Schober}, and H.~V. {Poor},
  ``{NOMA} assisted wireless caching: Strategies and performance analysis,''
  \emph{IEEE Transactions on Communications}, vol.~66, no.~10, pp. 4854--4876,
  2018.

\bibitem{pei2020tvt}
X.~{Pei}, H.~{Yu}, Y.~{Chen}, M.~{Wen}, and G.~{Chen}, ``Hybrid
  multicast/unicast design in {NOMA}-based vehicular caching system,''
  \emph{IEEE Transactions on Vehicular Technology}, vol.~69, no.~12, pp.
  16\,304--16\,308, 2020.

\bibitem{nomikos2021wc}
N.~{Nomikos}, T.~{Charalambous}, D.~{Vouyioukas}, and G.~K. {Karagiannidis},
  ``When buffer-aided relaying meets full duplex and {NOMA},'' \emph{IEEE
  Wireless Communications}, vol.~28, no.~1, pp. 68--73, 2021.

\bibitem{li2020tcom}
J.~{Li}, X.~{Lei}, P.~D. {Diamantoulakis}, F.~{Zhou}, P.~{Sarigiannidis}, and
  G.~K. {Karagiannidis}, ``Resource allocation in buffer-aided cooperative
  non-orthogonal multiple access systems,'' \emph{IEEE Transactions on
  Communications}, vol.~68, no.~12, pp. 7429--7445, 2020.

\bibitem{lei2021network}
L.~Lei, G.~Shen, L.~Zhang, and Z.~Li, ``Toward intelligent cooperation of {UAV}
  swarms: When machine learning meets digital twin,'' \emph{IEEE Network},
  vol.~35, no.~1, pp. 386--392, 2021.

\bibitem{lahmeri2021ojcoms}
M.-A. Lahmeri, M.~A. Kishk, and M.-S. Alouini, ``Artificial intelligence for
  {UAV}-enabled wireless networks: A survey,'' \emph{IEEE Open Journal of the
  Communications Society}, vol.~2, pp. 1015--1040, 2021.

\bibitem{qu2021network}
Y.~Qu, C.~Dong, J.~Zheng, H.~Dai, F.~Wu, S.~Guo, and A.~Anpalagan, ``Empowering
  edge intelligence by air-ground integrated federated learning,'' \emph{IEEE
  Network}, vol.~35, no.~5, pp. 34--41, 2021.

\bibitem{challita2019wcm}
U.~Challita, A.~Ferdowsi, M.~Chen, and W.~Saad, ``Machine learning for wireless
  connectivity and security of cellular-connected {UAV}s,'' \emph{IEEE Wireless
  Communications}, vol.~26, no.~1, pp. 28--35, 2019.

\bibitem{yang2020twc}
K.~Yang, T.~Jiang, Y.~Shi, and Z.~Ding, ``Federated learning via over-the-air
  computation,'' \emph{IEEE Transactions on Wireless Communications}, vol.~19,
  no.~3, pp. 2022--2035, 2020.

\bibitem{sun2019wcm}
X.~Sun, D.~W.~K. Ng, Z.~Ding, Y.~Xu, and Z.~Zhong, ``Physical layer security in
  {UAV} systems: Challenges and opportunities,'' \emph{IEEE Wireless
  Communications}, vol.~26, no.~5, pp. 40--47, 2019.

\bibitem{wang2019wcm}
H.-M. Wang, X.~Zhang, and J.-C. Jiang, ``{UAV}-involved wireless physical-layer
  secure communications: {O}verview and research directions,'' \emph{IEEE
  Wireless Communications}, vol.~26, no.~5, pp. 32--39, 2019.

\bibitem{zhang2020tvt}
Y.~Zhang, Z.~Mou, F.~Gao, J.~Jiang, R.~Ding, and Z.~Han, ``{UAV}-enabled secure
  communications by multi-agent deep reinforcement learning,'' \emph{IEEE
  Transactions on Vehicular Technology}, vol.~69, no.~10, pp. 11\,599--11\,611,
  2020.

\bibitem{nomikos2022access}
N.~Nomikos, S.~Zoupanos, T.~Charalambous, and I.~Krikidis, ``A survey on
  reinforcement learning-aided caching in heterogeneous mobile edge networks,''
  \emph{IEEE Access}, vol.~10, pp. 4380--4413, 2022.

\bibitem{liu2022ojcoms}
Q.~Liu, H.~Liang, R.~Luo, and Q.~Liu, ``Energy-efficiency computation
  offloading strategy in {UAV} aided {V2X} network with integrated sensing and
  communication,'' \emph{IEEE Open Journal of the Communications Society},
  vol.~3, pp. 1337--1346, 2022.

\bibitem{tran2022ojvt}
D.-H. Tran, S.~Chatzinotas, and B.~Ottersten, ``Satellite- and cache-assisted
  {UAV}: {A} joint cache placement, resource allocation, and trajectory
  optimization for {6G} aerial networks,'' \emph{IEEE Open Journal of Vehicular
  Technology}, vol.~3, pp. 40--54, 2022.

\bibitem{wu2022arxiv}
\BIBentryALTinterwordspacing
W.~Wu, F.~Zhou, B.~Wang, Q.~Wu, C.~Dong, and R.~Q. Hu, ``Unmanned aerial
  vehicle swarm-enabled edge computing: Potentials, promising technologies, and
  challenges,'' \emph{CoRR}, vol. abs/2201.08517, 2022. [Online]. Available:
  \url{https://arxiv.org/abs/2201.08517}
\BIBentrySTDinterwordspacing

\end{thebibliography}
\let\addcontentsline\oldaddcontentsline

\tocless{\begin{IEEEbiographynophoto}{Nikolaos Nomikos} (Senior Member, IEEE) received the Diploma
degree in electrical engineering and computer technology from the University of Patras, Greece, in 2009, and the M.Sc. and Ph.D. degrees from the Information and Communication Systems Engineering Department,
University of the Aegean, Samos, Greece, in 2011 and 2014, respectively. He is currently a Senior Researcher at the Department of Port Management and Shipping, National and Kapodistrian University of Athens. His research interest is focused on cooperative communications, non-orthogonal multiple access, full-duplex communications, and machine learning for wireless networks optimization. Dr. Nomikos is an Associate Editor of Frontiers in Communications and Networks.  He is a member of the IEEE Communications Society and the Technical Chamber of Greece.
\end{IEEEbiographynophoto}}

\tocless{\begin{IEEEbiographynophoto}{Panagiotis K. Gkonis} received the Diploma
and Ph.D. degrees in electrical and computer engineering and the M.Sc. degree in
engineering-economic systems programme from
the School of National Technical University of
Athens (NTUA), Greece, in 2005, 2009, and 2009,
respectively. From 2010 to 2015, he worked as a
Postdoctoral Researcher at the Intelligent Communications and Broadband Networks Laboratory
(ICBNET), NTUA. Moreover, he has also served
as a Scientific/Laboratory Associate for the Departments of Electrical and
Aircraft Engineering, (former) Technological Institute of Sterea Ellada.
In December 2015, he was appointed as an Electrical Engineer at Hellenic
Electricity Distribution Network Operator (HEDNO). He is currently an
Assistant Professor with the Department of Digital Industry Technologies, National and Kapodistrian University of Athens (NKUA). He has
authored/coauthored more than 60 scientific publications in international
journals, scientific conferences, and book chapters in the areas of wireless
networks, broadband communications, and computational modeling of cellular networks. As a Researcher, he has participated in various national and
European Funded Projects.
\end{IEEEbiographynophoto}}

\tocless{\begin{IEEEbiographynophoto}{Petros S. Bithas} (Senior Member, IEEE) received the Diploma (5 years) in electrical and computer engineering and PhD degree, with specialization in “Wireless Communication Systems”, from the University of Patras, Greece, in 2003 and 2009, respectively. During 2009-2018 he was affiliated with the Department of Electronics Engineering of the Technological Educational Institute of Piraeus, Greece. During 2010-2021, he was an associate researcher at the Department of Digital Systems, University of Piraeus (UNIPI), Greece. He is currently an assistant Professor at the Department of Digital Industry Technologies of the National \& Kapodistrian University of Athens, Greece. Dr Bithas serves on the Editorial Board of the International Journal of Electronics and Communications (ELSEVIER), Telecom, Drone (MDPI) and Wireless Communications and Mobile Computing (Hindawi). He has been selected as an “Exemplary Reviewer” of IEEE COMMUNICATIONS LETTERS and IEEE TRANSACTIONS ON COMMUNICATIONS in 2010 and 2020-2021, respectively. He has published more than 50 articles in International scientific journals and 40 articles in the proceedings of International conferences. His current research interests include stochastic modeling of wireless communication channels as well as design and performance analysis of vehicular communication systems.\end{IEEEbiographynophoto}}

\tocless{\begin{IEEEbiographynophoto}{Panagiotis Trakadas} received the Dipl.-
Ing. degree in electrical and computer engineering
and the Ph.D. degree from the National Technical University of Athens (NTUA). In the past,
he was worked at Hellenic Aerospace Industry
(HAI), as a Senior Engineer, on the design of military wireless telecommunications systems, and the
Hellenic Authority for Communications Security
and Privacy, where he was holding the position of
the Director of the Division for the Assurance of
Infrastructures and Telecommunications Services Privacy. He is currently
an Associate Professor with the National and Kapodistrian University of
Athens. He has been actively involved in many EU FP7 and H2020 Research
Projects. He has published more than 130 papers in magazines, journals, and
conference proceedings. His research interests include the fields of wireless
and mobile communications, wireless sensor networking, network function
virtualization, and cloud computing. He is a Reviewer in several journals,
including IEEE TRANSACTIONS ON COMMUNICATIONS and IEEE TRANSACTIONS
ON ELECTROMAGNETIC COMPATIBILITY journals.
\end{IEEEbiographynophoto}}

\end{document}